\input phyzzx
%
%







\def\a{\alpha}

\def\b{\beta}

\def\s{\sigma}


\def\o{\over}

\def\bold#1{\setbox0=\hbox{$#1$}
     \kern-.025em\copy0\kern-\wd0
     \kern.05em\copy0\kern-\wd0
     \kern-.025em\raise.0433em\box0 }
\def\lowmp{\lower.11em\hbox{${\scriptstyle\mp}$}}

\def\frac#1#2{{\textstyle{
 #1 \over #2 }}}                            


\def\1{{\rm 1 \!\!\, l}}                        
\def\sca#1#2{\left( #1 \, , \, #2 \right)} 
%

%
%


\hyphenation{Di-par-ti-men-to}
\hyphenation{na-me-ly}
\hyphenation{al-go-ri-thm}
\hyphenation{pre-ci-sion}
\hyphenation{cal-cu-la-ted}

%

%
%
%
\Pubnum={PAR-LPTHE-93/43}\date{August, 1993}
\hyphenation{Schwarz-schild}
\chardef\oo="1C
%
%
\def\LPTHE{\address{Laboratoire de Physique Th\'eorique et Hautes \'Energies
\break
Universit\'es Paris VI-VII - Laboratoire associ\'e au CNRS n$^{\rm o}$ 280
\break
Tour 16, 1er. \'et., 4, Place Jussieu, 75252 Paris Cedex 05
\break
FRANCE}}

%
%
\titlepage
\title{Strings in cosmological and black hole backgrounds: ring solutions}
\author{H.J. de Vega and 
I.L. Egusquiza }
\LPTHE
\vskip15pt
\abstract{
The string equations of motion and constraints are solved
for a ring shaped Ansatz in cosmological and black hole spacetimes.
In FRW universes with arbitrary power  behavior [$R(X^0) =
a\;|X^0|^{\a}\, $], the asymptotic form of the solution is found for
 both $X^0 \to 0$ and $X^0 \to \infty$ and we plot the numerical
solution for all times. Right after the big bang ($X^0 = 0$)
, the string energy
decreasess as $ R(X^0)^{-1} $ and the string size grows as $ R(X^0) $ 
for $ 0 < \a < 1 $ and as $ X^0 $ for $ \a < 0 $ and $ \a > 1 $.
Very soon [ $ X^0 \sim 1 $] , the ring reaches its oscillatory regime
with frequency equal to the winding and  constant size and energy.
This picture holds for all values of $ \a $ including string
vacua (for which, asymptotically, $ \a = 1$). In addition,
an exact non-oscillatory ring solution is found.

For black hole spacetimes (Schwarzschild, Reissner-Nordstr\oo m and
stringy), we solve for ring strings moving
towards the center. Depending on their initial conditions
(essentially the  oscillation phase), they are are absorbed or not
by Schwarzschild black holes. The phenomenon of particle transmutation is
explicitly observed (for rings not swallowed by the hole). An
effective horizon is noticed for the rings. Exact and explicit ring
solutions inside the horizon(s) are found. They may be interpreted as
strings propagating between the different universes described by the
full black hole manifold.

  }
\endpage
\REF\dVS{H.J.~de~Vega and
N.~S\'anchez \journal Phys.Lett.&B197 (87) 320.}
\REF\dVer{H.J.~de~Vega and N. S\'anchez, Lectures delivered
at the Erice School ``String Quantum Gravity and Physics at the Planck
scale'', 21-28 June 1992, Proceedings edited by N. S\'anchez,
World Scientific, 1993.}
\REF\twbk{See for a review, T.W.B. Kibble, Erice Lectures at the
Chalonge School on Astrofundamental Physics,
 N. S\'anchez editor, World Scientific, 1992.} 
\REF\dNu{H. J. de Vega and N. S\'anchez \journal
Phys. Rev. &D45 (92) 2783.}
\REF\dRSu{H. J.~de~Vega, M.~Ram\'on~Medrano and N.~S\'anchez, LPTHE  Paris
  preprint  92-13. To appear in Classical and Quantum Gravity.}
\REF\dNc{H. J. de Vega and N. S\'anchez\journal 
Nucl. Phys. &B317 (89) 706. \hfill \break
D. Amati and K. Klim\v{c}ik\journal Phys. Lett. &B210 (88) 92.}
\REF\CdV{M. Costa and H. J. de Vega 
\journal Ann. Phys. &211, (91) 223 and 235.}
\REF\dRS{H.J.~de~Vega, M.~Ram\'on~Medrano and N.~S\'anchez
\journal Nucl.Phys.&B374 (91) 405.}
\REF\dN{H.J.~de~Vega and
N.~S\'anchez \journal Phys.Rev.&D42 (90) 3969.}
\REF\agn{H. J. ~de ~Vega and N. S\'anchez \journal 
Nucl. Phys. &B309 (88) 552 and 577 . \hfill \break
C. Loust\'o and N. S\'anchez \journal  Phys. Rev. &D45 (93) 4498 .}
\REF\SV{N.~S\'anchez and
G.~Veneziano \journal Nucl.Phys.&B333 (90) 253.}
\REF\GSV{M.~Gasperini, N.~S\'anchez and G.~Veneziano 
\journal Int.J.Mod.Phys.&A6 (91) 3853.}
\REF\tras{H.J.~de~Vega, M.~Ram\'on~Medrano and N.~S\'anchez
\journal Nucl.Phys.&B351 (91) 277.}
\REF\horowitz{G.T.~Horowitz, 
to appear in the proceedings of the 1992 Trieste Spring School, 
Santa Barbara preprint UCSBTH-92-32, hep-th/9210119.}
\REF\VSexact{H.J.~de~Vega and N.~S\'anchez 
\journal Phys.Rev.&D47 (93) 3394.}
\REF\dVMS{H.J.~de~Vega, A.V.~Mikhailov
and N.~S\'anchez, ``Exact string solutions in 2+1 dimensional de Sitter
spacetime'' , lpthe preprint 92/32, hep-th/92-09-047.}
\REF\dVMSd{F. Combes, H.J.~de~Vega, A.V.~Mikhailov and N.~S\'anchez,
lpthe preprint 93/44. } 
\REF\vil{ A. ~Vilenkin \journal Phys. Rev. &D24 (81) 2082
\journal Phys. Rep. &121 (85) 263.\hfill \break
N. ~Turok and P. ~Bhattacharjee \journal Phys. Rev. &D29 (84) 1557.}
\REF\tseyt{A.A.~Tseytlin in the Proceedings
of the Erice School ``String Quantum Gravity and Physics 
at the Planck Energy Scale'', 21-28 June 1992, 
Edited by N. ~S\'anchez, World Scientific, 1993. }
\REF\blackb{``Gravitation'', C. W. Misne, K. S. Thorne and 
J. A. Wheeler, W. H. Freeman, 1973.}
\REF\mate{S.~Wolfram, ``Mathematica: a system for doing mathematics by 
computer.'', 2nd edition (1991), Addison-Wesley Pub. Co., California, USA.}\REF\myers{R.~Myers \journal Phys. Lett. & B199 (87) 371.}
\REF\mueller{M.~Mueller
\journal Nucl. Phys. & B337 (90) 37.}
\REF\lm{S.~ Lonsdale and I.~ Moss \journal Nucl. Phys.&B298 (88) 693.}
\REF\fal{V. P. Frolov et al \journal Phys. Lett. &B224 (89) 255. }
\REF\lar{A.~L.~Larsen \journal Phys. Lett. &B273 (91) 375 and
Nordita preprint (in preparation)}
\REF\fl{A. L. Larsen and V. Frolov, Nordita preprint, 93/17.\hfill \break
A. L. Larsen, Nordita preprint, 93/52.}
\REF\horo{J.~Horne and G.~Horowitz, ``Cosmic Censorship and the Dilaton'',
preprint Yale (YCTP-P17-93) - ITP Santa Barbara (NSF-ITP-93-95); 
hep-th/9307177.}
\REF\garymae{G.~Gibbons and K.~Maeda \journal Nucl.Phys.&B298 (88) 741.}
\FIG\figkdneg{Cosmological spacetime with $k=2$: numerical solution to
the ringlike equations of motion for 
initial conditions such that $\eta(0)\dot\eta(0)<0$.}
\FIG\figkcneg{Cosmological spacetime with $k=4$: numerical solution,
for initial conditions
such that $\eta(0)\dot\eta(0)<0$.}
\FIG\figkdpos{Cosmological spacetime with $k=2$: initial conditions
such that $\eta(0)\dot\eta(0)>0$.}
\FIG\figkcpos{Cosmological spacetime with $k=4$: initial conditions
such that $\eta(0)\dot\eta(0)>0$.}
\FIG\figkdtot{Cosmological spacetime with $k=2$: both asymptotic regimes
are present here.}
\FIG\figktra{Cosmological spacetime with $k=-3/2$: close to the vanishing 
point for $\eta$. Note the analyticy of the solution.}
\FIG\figktrb{Cosmological spacetime with $k=-3/2$: The
large $\tau$ asymptotic regime.}
\FIG\figmyers{Myers' spacetime: both asymptotics are
observed.}
\FIG\figmuellera{Mueller's spacetime: the large $\tau$ asymptotic behaviour
is clearly observed in this numerical solution.}
\FIG\figmuellerb{Mueller's spacetime: A closer look to the limit
$X^0\to1^+$.}
\FIG\figbhint{Exact solution to the equations of motion inside the
horizon of a Schwarzschild black hole, in Kruskal-Szekers coordinates:
the string worldsheet covers all the equatorial section of the interior
of the horizon.}
\FIG\figbhfal{Numerical solution for the equations of motion of a string 
in a Schwarzschild black hole background: the string falls into the
black hole.}
\FIG\figbhpas{Numerical solution for the equations of motion of a string 
in a Schwarzschild black hole background: the string falls into the
black hole, but only after first going past it and then back into
the singularity.}
\FIG\figbhreb{Numerical solution for the equations of motion of a string 
in a Schwarzschild black hole background: the string goes past the black
hole, circles round it, and then bounces back with
a change in its amplitude and momentum.}

\chapter{Introduction}
 The systematic investigation of string dynamics  in curved spacetimes
started in [\dVS ] has shown a variety of new physical phenomena
[\dVer ]. These results  are relevant both for fundamental (quantum)
 strings and for cosmic strings, which behave in an  essentially
classical  way [\twbk].

String propagation has been studied in non-linear gravi\-tational
plane waves [\dNu,\dRSu] and shock-waves 
[\dNc,\CdV ], conical spacetimes
[\dRS,\dN], black holes [\agn ], and cosmological spacetimes [\dVer,\SV,\GSV ].

Among the cosmological backgrounds, de Sitter occupies a special
place. On one hand, it is the relevant inflationary geometry, and on the
other, string propagation turns out to be special there [\dVS , 
\SV ,\GSV ]. 
Moreover the classical string equations of motion (plus the
string constraints) happen to be integrable in D-dimensional de Sitter 
universe 
[\VSexact , \dVMS,\dVMSd ]. More precisely, they are equivalent to a
sigma model on the grassmanian $SO(D,1)/O(D)$ with periodic
boundary conditions (for closed strings) or  Neumann boundary
conditions (for open strings).

For generic Friedmann-Robertson-Walker (FRW) cosmological spacetimes,
the propagation of strings is certainly a non-integrable
problem. Therefore, one is faced  with a rather
formidable set of coupled non-linear partial differential
equations. Analogous difficulties arise in other non-integrable
backgrounds, such as Schwarszchild  black holes.
To grasp basic physical properties,
we consider in the present paper solutions for the
motion of classical closed test strings for which the $\tau$
and $\sigma$ dependence are separated. (Hence, we have to solve
just non-linear ordinary differential equations).
That is, we make separable ring-shaped ans\"atze sharing 
the symmetry of the background geometry. 

As is well known,  the effective action for string
theory corresponds to a modification of Einstein-Hilbert's action
(see for instance the review [\tseyt] for the
tree-level effective action). This leads to cosmological
spacetimes that are solutions of the variational equations for
this action. It is therefore interesting to
investigate ring-like solutions in these
stringy cosmological spacetimes as well, and to compare their features 
 with those encountered in our study of FRW spacetimes

We consider cosmological spacetimes with metric
$$
{\rm d}s^2 =({\rm d}X^0)^2-R(X^0)^2 \;\sum_{i=1}^{D-1} ({\rm d}X^i)^2
\eqn\metc
$$
For FRW universes $ R(X^0) = a ~ |X^0|^{\alpha} $ , where $a$ is a
constant , $\alpha$ is 
$1/2$ for radiation dominated universes and $2/3$ for matter dominated
universes. However, we shall consider arbitrary real values of  $\alpha$.
In particular, for  $\alpha = 1$ we have a tree level string vacuum [\myers].
String vacua may  give more complicated functions $ R(X^0) $,
but $ R(X^0) $ usually grows as $X^0$ for large $X^0$ [\tseyt ].

The ring-shaped ansatz for the string solution is defined by
$$\eqalign{
X^0&=X^0(\tau)\,,\cr
X^1&=f(\tau)\cos\sigma\,,\cr
X^2&=f(\tau)\sin\sigma\,,\cr
X^i&={\rm const.}\,,\quad i\geq3\,.\cr}\eqn\anillo$$
and leads to two coupled ordinary nonlinear differential
equations on two functions : the ring  radius $  f(\tau) $ (in comoving
coordinates) and the cosmic time $X^0 = X^0(\tau)$.
The other relevant physical quantities here are the invariant string size
$ S(\tau) \equiv f(\tau) \; R(X^0(\tau))\;$ and the string energy
$ E(X^0) = {1 \o {\a'}} \; {\dot X}^0(\tau) $. Here $\tau$ stands for the
proper string time and $\a'$ for the string tension .

We can summarize our results as follows for all power expanding  universes 
[$ R(X^0)\allowbreak \buildrel{{X^0\to+\infty}}\over= a \;
 |X^0|^{\alpha} $].
At the time $X^0 = 0$ singularity (big bang/big crunch), 
the string energy is infinite
and its size is zero. Right after that, the energy redshifts as:
$$
E(X^0)\buildrel{{X^0\to 0}}\over\sim {\rm const}. {1 \o {\a' R(X^0)}}
\eqn\enas
$$
and the invariant size grows as follows
$$\eqalign{
S = {\rm cte.} ~ R(X^0) \quad {\rm for} \quad  X^0 \to 0 \quad {\rm
when } \quad 0 <  \alpha < 1  \cr
S = {\rm cte.} ~ X^0  \quad {\rm for} \quad X^0  \to 0 \quad {\rm when}
  \quad \a > 1 \quad {\rm and~~ when } \quad \alpha < 0 \cr  }
\eqn\colap
$$
Notice that in the second case the string size is proportional to the
particle horizon ($\sim X^0 $). In the first case, the string grows
as the expansion factor. This is exactly like the case of strings
falling into a singular plane wave [\dNu ,\dRSu ].
Stringy universes are special ($\alpha = 1$) and we find (sec. 3)
that 
$$
S = R(X^0) \log{\left[{1} \over {R(X^0)}\right]} 
 \quad {\rm for} \quad R(X^0) \to 0 .
\eqn\colaq
$$
Very soon [ $ X^0 \sim 1 $] , the string ring solution reaches 
an oscillatory regime with  [see eqs. (2.23), (2.33) and (3.4)]
$$
f(\tau) \buildrel{{X^0\to+\infty}}\over\sim C\;{1
\over {\tau^{\alpha}}}\;\cos(\tau + \varphi)
\eqn\osci
$$
with $X^0$ proportional to $ C^{1-\a} ~ \tau$. Here, $C$ and
$\varphi$ are arbitrary constants, amplitude and  phase, 
respectively. This describes for large
$X^0$ a string of constant invariant size that oscillates with unit
frequency. Quantum mechanically, eq.\osci\ corresponds to an
excitation  of graviton or dilaton type. 
Multiple winded ring-shaped strings oscillate
with arbitrary integer frequency (see eq.(2.33) ). 
Notice that simple string oscillatory behaviors like \osci\
are absent in inflationary cosmologies as the de Sitter universe.
 
In conclusion, when the universe expands as any finite power of
$X^0$ , the ring string oscillates similarly to flat
spacetime. Only the oscillation amplitude varies with time
in such a way that the invariant size, $S(\tau)$, 
and the energy remain constant.
Near the big bang (or big crunch) singularities
the string collapses (eqs \colap ,\colaq ).

We find in addition an exact ring solution with radius proportional to
the conformal time: 
$$\eta(\tau)= K e^{\pm \tau/\sqrt{-k-1}}\,,\quad 
f(\tau)=  {K\over{{\sqrt{-k}}}}e^{\pm \tau/\sqrt{-k-1}}
\,.\eqn\aniex$$
where $K$ is an arbitrary constant. This solution is real for
$ k + 1 < 0 $ . For $ -1 < k < 0 $, it is real for imaginary
$\tau$ . For $ k > 0 $ , this solution is real for imaginary
conformal time $\eta$ and  imaginary string time $\tau$. Hence,
it can be considered as an {\it instanton} for $ k > - 1$.

Since Schwarzschild black holes are asymptotically flat spacetimes,
we consider these ring string solutions starting very far from 
the black hole. They asymptotically behave as Minkowski solutions with
a momentum $p$ directed towards the black hole center, an oscillation
amplitude $m$ and an energy $e$ with $e^2 = p^2 + m^2$.
This energy is in fact proportional to the string energy
at all times $t$, $E(t)=e/\a'$.
Notice that the oscillation amplitude coincides with the classical
string mass. For ring strings wounded $n$-times, the mass turns to
be $n m$. When the ring string approaches the black hole, its
oscillations do not qualitatively change, as can be seen from 
figs. 14. Then, the ring is swallowed or not by the black hole,
depending on how much the string approaches the
singularity.
There exists an effective horizon within which the ring string
is always absorbed by the black hole. We find that the sphere $r =  
(3/2) R_s $ (where $ R_s $ = Schwarzschild radius) is completely
inside this effective horizon. Numerical results indicate that the 
effective horizon here is close to the
effective horizon for massless geodesics  $ {3 \over 2} \sqrt3 \;  r_s 
= 2.5980...  R_s $.

Absorption by the black hole may occur for any value of the initial
amplitude and momentum. The oscillation phase determines whether the
ring will be swallowed or not. When the ring is not absorbed it
gets out after turning an even or odd number of $\pi's$ around the
center. For an even (odd) number we have back- (forward)
scattering. The outgoing oscillation amplitude, $m'$ is generically
different from the initial amplitude $m$. This is an explicit
illustration of the phenomenon of particle transmutation noticed
in ref.[\tras ].

Several black hole perturbative string vacua are now known (for a
recent review, see [\horowitz]). In particular there exist 
rotationally invariant solutions which are generalizations of 
Reissner-Nordstr\oo m's spacetime. 

We show how a generic rotationally
symmetric spacetime admits ringlike solutions to the string
equations of motion, and apply this to the aforementioned black
hole solutions. The analysis of the possiblity of collapse of the string onto
the singularity is radically different from the behaviour
present in Schwarzschild's background. Whenever the region
close to the singularity is static (as would be the case for
Reissner-Nordstr\oo m), the singularity is strongly repulsive
of the strings, similarly to what happens with point particles.
Only if the usually present inner horizon is absent or coincides
with the singularity will we get collapsing
string with the worldsheet lacking analyticity at the collapse point.
The asymptotic region $r>>|\dot\theta|$, which corresponds to the
asymptotically Minkowskian region of all these spacetimes,
does indeed reproduce the expected oscillatory behaviour.

An exact solution for the string equations of motion is obtained
for Reissner-Nordstr\oo m and stringy black holes.
It can be understood as the propagation of the string from
one asymptotically flat region to another, crossing both the
outer and the inner horizon present in these black holes.
This is clearly analogous to the
possibility that timelike curves can connect the different
asymptotically flat regions.

 The corresponding solution for Schwarzschild's background does
exist, but remains always inside the horizon. Thus no violation
of causality can be caused by the ringlike strings under consideration.

\chapter{Strings propagating in Friedman-Robertson-Walker
 cosmological spacetimes}

We consider in this section closed test strings propagating 
in FRW spacetimes with metric
$$\eqalign{
{\rm d}s^2&=({\rm d}X^0)^2-R(X^0)^2 \;\sum_{i=1}^{D-1} ({\rm d}X^i)^2=\cr
&=R(\eta)^2\left[({\rm d}\eta)^2 -\sum_{i=1}^{D-1} ({\rm d}X^i)^2\right]\,,
\cr}\eqn\metric$$
where $X^0$ stands for cosmic time and $\eta$ for conformal time. 
These times are related by
$$
\eta=\int^{X^0}{{{\rm d}X^0}\over{R(X^0)}}\,.
\eqn\tiem
$$
The function $R(\eta)^2$ will be assumed of power type:
$$
R(\eta)^2=A\;\eta^k\,,
\eqn\power
$$
where $A$ is a constant. For $k=-2$ we have de Sitter spacetime, 
for $k=2$  a radiation 
dominated FRW universe, and for $k=4$ a matter dominated universe 
[\blackb].

For $k \neq -2$ we get from eqs.\tiem -\power :
$$
X^0 = {{2\sqrt A}\over{k+2}}~\eta^{{k+2}\over2} \quad {\rm and}\quad
\alpha = {k \o {k+2}}
\eqn\tien
$$
The lagrangian function for a bosonic string in the geometry \metric\
takes the form
$${\cal L}={1\over2}\left[(\partial_\mu X^0)^2 - 
R(X^0)^2 \; (\partial_\mu X^i)^2\right]\,,\eqno\eq$$
where $\partial_\mu$ are derivatives w.r.t. the worldsheet coordinates
$x_\mu$, $x_0=\tau$ and $x_1=\sigma$ (we use the conformal
gauge throughout). This yields the equations of motion
$$\eqalign{
\partial^2X^0& - R(X^0)\; {{{\rm d}R}\over{\,{\rm d}X^0}}\;
\sum_{i=1}^{D-1}(\partial_\mu X^i)^2=0\,,\cr
\partial_\mu&\left[R^2 \partial^\mu X^i\right]=0\,,\qquad 1\leq i\leq D-1\,,\cr
}\eqn\movim$$
and the constraints
$$T_{\pm\pm}=(\partial_\pm X^0)^2 - R(X^0)^2 \; (\partial_\pm X^i)^2=0\,.
\eqn\constr$$
Equations \movim -\constr\ turn to be integrable for the de Sitter 
case ($k=-2$) [\VSexact].
They are not integrable for generic $k$. We shall therefore proceed to
analyze a simple ansatz where variables $\sigma$ and $\tau$ separate.
We choose a ring configuration whose radius depends on $\tau$:
$$\eqalign{
X^0&=X^0(\tau)\,,\cr
X^1&=f(\tau)\cos\sigma\,,\cr
X^2&=f(\tau)\sin\sigma\,,\cr
X^i&={\rm const.}\,,\quad i\geq3\,.\cr}\eqn\ansatz$$
We then find {}from eqs. \movim -\constr\ the following set of 
ordinary differential equations for $X^0(\tau)$ and $f(\tau)$:
$$\eqalign{
{{{\rm d}}\over{{\rm d}\tau}}
\left[R^2{{{\rm d}f}\over{{\rm d}\tau}}\right] + R^2 f&=0\,,\cr
{{{\rm d}^2 X^0}\over{{\rm d}\tau^2}}+ 
R {{{\rm d}R}\over{{\rm d}X^0}}\left(\dot f^2 - f^2\right)&=0\,,\cr
\left({{{\rm d}X^0}\over{{\rm d}\tau}}\right)^2 
- R^2\left(\dot f^2 + f^2\right)&=0\,,\cr}\eqn\deru$$
where $\dot{f} $ stands for ${{{\rm d}f}\over{{\rm d}\tau}}$.

The string energy can be easily computed from the spacetime
string energy-momentum tensor:
$$
\sqrt{-G}~ T^{AB}(X) = {1 \o {2\pi \a'}} \int d\s d\tau
\left( {\dot X}^A {\dot X}^B -X'^A X'^B \right) \delta^{(D)}(X - X(\s, \tau) )
\eqn\tens
$$
Therefore, whenever $X^0= X^0(\tau)$, the string energy at a time 
$X^0$ is given by:
$$
E(X^0) = \int d^{D-1}X \sqrt{-G}~ T^{00}(X) = {1 \o {\a'}}
{{dX^0} \o {d\tau}}
\eqn\eneg
$$ 

For the restriction to power-type metrics, eq.\power, 
eqs.\deru\ take a particularly simple form for the conformal time
$\eta=\eta(\tau)$:
$$\eqalignno{
\eta\ddot\eta + k \dot f^2 &=0\,,&\eqnalign{\andu}\cr
\eta\ddot f + k\dot\eta\dot f + \eta f&=0\,&\eqnalign{\andd}\cr
\dot\eta^2-\dot f^2 - f^2&=0\,.&\eqnalign{\andt}\cr}$$
The last equation is due to the string constraint. Notice that 
the first two equations guarantee that $[\dot\eta^2-\dot f^2 - f^2]$
is a constant of the motion. Hence the constraint \andt\ will
be satisfied for all times if it is satisfied by the initial
conditions. We have therefore only two independent differential
equations in the set \andu-\andt, with one constraint on the
possible initial conditions. In addition to this, the equations
are invariant under the scaling transformation
$$\eta(\tau)\to\lambda\eta(\tau)\,,\quad f(\tau)\to\lambda f(\tau)\,,
\eqn\sca$$
where $\lambda$ is an arbitrary constant. It thus follows that
the space of solutions is parametrised by two initial conditions
and a scaling factor.

The invariant spacetime interval \metric\ measured for these string 
solutions takes the form
$${\rm d}s^2=\eta^k f^2\left[{\rm d}\tau^2-{\rm d}\sigma^2\right]\,,\eqno\eq$$
where we have used eq. \andt. We can then interpret 
$$S(\tau)^2 = \eta^k f^2\eqn\siz$$
as the string size squared [\dVer].

Eqs. \andu-\andt\ admit a pair of simple solutions that can
be written in closed form:
$$\eta(\tau)=e^{\pm \tau/\sqrt{-k-1}}\,,\quad 
f(\tau)={1\over{{\sqrt{-k}}}}e^{\pm \tau/\sqrt{-k-1}}
\,.\eqn\expo$$
Of course, these two solutions can be scaled according to \sca. 
They are real for $k+1<0$, and describe a string that explodes
or collapses for large $\tau$, which can be seen {}from the form
taken in this case by the string size $S(\tau)$,
$$S(\tau)^2 =-{1\over{k}}\exp\left(\pm{{k+2}\over{{\sqrt{-k-1}}}}\tau\right)\,.
\eqno\eq$$
Clearly, the string explodes, for $\tau\to+\infty$,
for the choice of plus sign of the
exponent, $k+2$ being larger than zero ($-1>k>-2$), or the choice of
the minus sign whenever $k+2<0$.
The converse situation, i.e., string collapse for $\tau\to+\infty$, 
occurs for the choice of the minus sign  and $k+2>0$ ($-1>k>-2$), 
or plus sign
and $k+2<0$. Notice that for these solutions
$${{S(\tau)}\over{R(\tau)}}={1\over{\sqrt{-k}}}
\exp\left(\pm{{\tau}\over{\sqrt{-k-1}}}\right)\,,\eqno\eq$$
giving exponential collapse/explosion of the string with respect
to the evolution of the spacetime.

 Solution \expo\  is real for
$ k + 1 < 0 $ . For $ -1 < k < 0 $, it is real for imaginary
$\tau$ . For $ k > 0 $ , this solution is real for imaginary
conformal time $\eta$ and string time $\tau$. Hence,
it can be considered as a string  {\it instanton} for $ k > - 1$.
For $ k > 0 $ , it is a real solution in a universe
with euclidean signature.

The case $k=-2$ (de Sitter) is special. There, eq.\expo\ yields 
the solution $q_0(\tau)$ found in reference [\dVMS ].
In this case the string size remains constant.

Notice that \expo\ (for $k\neq-2$) collapses or explodes irrespective of
the initial string size. On the other hand, for de Sitter space, the 
string equations of motion and of constraint reduce to a sinh-Gordon
equation with potential $-2\cosh\alpha$ [\VSexact ], where the string size
squared is $S(\tau)^2 =\exp(\alpha(\sigma,\tau))/(2H^2)$. Therefore,
configurations with positive (resp. negative) $\alpha$ are driven 
towards $+\infty$ (resp. $-\infty$) $\alpha$. That is, strings such
that for any given initial moment they are of size bigger
(resp. smaller) than the horizon
($1/\sqrt2 H$) tend to explode (resp. collapse).

The peculiarity of de Sitter spacetime in this context can be clearly seen 
as soon as one realizes that eqs.\andu-\andt\ can be cast in a
hamiltonian form with a constraint.
Define 
$$H={1\over2} \eta^{-k}\left(\Pi_\eta^2-\Pi_f^2-\eta^{2k}f^2\right)\,,
\eqno\eq$$
and Poisson commutators the canonical ones,
$$\{\eta,\Pi_\eta\}=\{f,\Pi_f\}=1\,,$$
the rest of commutators between $f, \eta, \Pi_f, \Pi_\eta$ 
being equal to zero.

The equations of motion previously written down are then obtained {}from
$\dot g=\{g,H\}$ for any $g$ in the algebra $\cal A$ generated by
$f, \eta, \Pi_f, \Pi_\eta$.  The equation of constraint is 
equivalent to the dynamical constraint $H\approx0$.

We can write the following element $Q\in{\cal A}$,
$Q=\eta\Pi_\eta+f\Pi_f$. It will be the generator of dilations
of the string size.
Its time derivative is easily computed to be
$$\{Q,H\}=(k+2)\left(H+\eta^k f^2\right)=(k+2)\left(H+S^2\right)\,,\eqno\eq$$
whence for $k=-2$ we have an additional (quadratic) conserved quantity.
Notice that $S^2$ itself is not a conserved quantity in general, even for
de Sitter spacetime, since $\{S^2,H\}=kf^2\eta^{-1}\Pi_\eta-2f\Pi_f$.

The existence of this additional conserved quantity for this ansatz and
de Sitter spacetime reflects the fact that the motion of classical
test strings in de Sitter spacetimes is integrable [\VSexact]-[\dVMS].

Let us now derive the asymptotic behaviour of $\eta$ and $f$ analytically
{}from \andu-\andt, for solutions other than the previously
written exponentials. 

We find, for $\tau\to+\infty$ (and $k\neq-2$)
$$\eqalign{
\eta(\tau)&\buildrel{{\tau\to+\infty}}\over\sim \tau^{2/(k+2)}\,,\cr
X^0(\tau)&\buildrel{{\tau\to+\infty}}\over\sim 
{{2\sqrt{A}}\o{(k+2)}}\;\tau ~, \cr
f(\tau)&\buildrel{{\tau\to+\infty}}\over\sim {2\over{(k+2)}} \tau^{-k/(k+2)} 
\cos(\tau+\varphi)\,,\cr}
\eqn\solasim
$$
where $\varphi$ is a constant phase and oscillation amplitude has been
 normalized using the scaling \sca. 
The size of the set of solutions with this asymptotic behaviour
is asymptotically a constant times an oscillating term:
$$S(\tau)\buildrel{{\tau\to+\infty}}\over\sim \left({2\over{k+2}}\right)
|\cos(\tau+\varphi)|\,\buildrel{{\tau\to+\infty}}\over\sim {\sqrt2 \over
{(k+2)}}\;.
\eqn\sizasim$$
That is, \solasim\ describes the asymptotic behaviour of a string
whose size oscillates with unit frequency. Quantum mechanically,
this corresponds to an excitation of graviton or dilaton type.
Notice that this behaviour holds for all $k\neq-2$.

The result \solasim\ holds for large $R$ and is therefore valid for
any universe where eq.\power\ is valid {\it asymptotically} .
We want to stress that all cosmological geometries exhibit
simple oscillatory string behaviour for large $R$ except when $R(X^0)$
grows faster than any power of $X^0$. That is, simple oscillatory 
string behaviour
 does not appear in inflationary  universes like de Sitter.

The contraction and dilation of the universe in this limit, $\tau\to
\infty$, is now governed by $k$: for $-2<k<0$ this corresponds to a 
contracting universe; otherwise it will be expanding.

In summary, the asymptotic behaviour of the string given by 
\solasim\ is similar to the usual $|n|=1$ modes in Minkowski spacetime.
It corresponds to ``stable behaviour'' as defined in [\GSV] 
(the special case $ k=-2 $ (de Sitter spacetime) is analyzed in 
reference [\dVMS]).

Behaviours of the type \solasim\ for $\a > 0$ were called
``string stretching'' in refs.[\vil] referring to the fact that the
string amplitude $f(\tau)$ stretches when the universe expands. This
is actually a coordinate-dependent effect since the invariant string
size stays constant [eq. \sizasim ]. We prefer to use the term
'string stretching' in de Sitter universe where the {\bf invariant} string
size in a whole class of string solutions grows as fast as the
universe [\dVS ,\dVMS ,\dVMSd ].

Before describing our numerical study of eqs \andu-\andt, let
us consider the singular points where $\eta$ vanishes. Assuming
that $\eta(\tau_0)=0$, we find for $k>0$
$$\eqalign{
\eta(\tau)&\buildrel{\tau\to\tau_0}\over\sim (\tau-\tau_0)^{1/(k+1)}\,,\cr
f(\tau)&\buildrel{\tau\to\tau_0}\over\sim f_0\pm (\tau-\tau_0)^{1/(k+1)}\,.\cr}
\eqn\singu$$
For the case $0>k>-1$, 
$f_0$ must be set equal to $0$.
Expression \singu\ also holds as an asymptotic behaviour for $k<-1$ as 
$\tau\to\tau_0$, but in such a case, both $\eta(\tau)$ and
$f(\tau)$ diverge at $\tau=\tau_0$.
We have again made use of the scaling freedom given by \sca. The string 
size squared results to be
$$\eqalign{
S(\tau)^2 & \buildrel{\tau\to\tau_0}\over\sim
(\tau-\tau_0)^{(k+2)/(k+1)}\,\qquad {\rm for~}k<0\,;\cr
S(\tau)^2 & \buildrel{\tau\to\tau_0}\over\sim f_0^2 \;
(\tau-\tau_0)^{k/(k+1)}\,\qquad {\rm for~}k>0
\,.\cr}\eqno\eq$$
In this regime, the universe contracts (big crunch) as
$$R(\tau)^2\buildrel{\tau\to\tau_0}\over\sim (\tau-\tau_0)^{k/(k+1)}\to0
\,,\eqno\eq$$
whereas the string collapses for $k>0$ or $k<-2$, but explodes
for $-2<k<-1$.

Notice, that the string size $S(\tau)$ grows as $X_0(\tau)$ grows
for $k < 0$. That is, the string size is proportional to the
particle horizon in this regime. For $k>0$, the string size behaves
here as
$R(X_0(\tau))$.

{F}rom eqs.\eneg -\singu\ we find that the string energy $ E(X^0) =
{1 \o {\a'}}~{\dot X}^0 $
behaves near the big-bang ($X^0 = 0$) singularity as
$$
E(X^0) \sim {{\rm const.} \o {\a'}} \tau^{-{k \o {k+1}}} \sim
 {{\rm const.} \o {\a'}} {1\o {R(X^0)}}
\eqn\ensin
$$
Notice that $E(X^0)$ decreases for growing small $X^0$ with the
gravitational red-shift factor $1/R(X^0)$ . On the contrary, when the
particle horizon ($\sim X^0 $) is much larger than the ring size,
the string energy is no more redshifted and tends to a constant
(eq. \solasim ). 

Another possible behaviour is given by
$$\eqalign{
\eta(\tau)&\buildrel{\tau\to\tau_0}\over\sim \tau-\tau_0\,,\cr
f(\tau)&\buildrel{\tau\to\tau_0}\over\sim 1 - 
{{(\tau-\tau_0)^2}\over{2(k+1)}}\,.\cr}\eqn\exun$$
For $0>k>-1$, a more likely behaviour is given
by
$$\eqalign{
\eta(\tau)&\buildrel{\tau\to\tau_0}\over\sim \tau-\tau_0\,,\cr
f(\tau)&\buildrel{\tau\to\tau_0}\over\sim 1 + a
(\tau-\tau_0)^{1-k}\,,\cr}\eqn\exdos$$
with $a$ a constant.
Both sets of equations correspond to a collapsing (resp. exploding) string for
$k>0$ (resp. $k<0$), with size
$S(\tau)\buildrel{\tau\to\tau_0}\over\sim (\tau-\tau_0)^{k/2}$.
The behaviour \exun\ is however subdominant when compared to \singu\ for
$k>0$. For $k<-1$ \singu\ is singular, and, since, as we shall later prove,
$\eta$ must vanish for some finite value of $\tau$, it follows that 
\exun\  will be present. Similarly for $-1<k<0$.

The equations of motion \andu-\andt\ being invariant under
time reversal ($\tau\to-\tau$), both asymptotic behaviours,
\solasim\ and \singu, may describe initial or final situations
of the string. The numerical analysis, carried out using
Mathematica [\mate], precisely shows how
to connect such behaviours.

Equation \andu\ tells us that $\eta\ddot\eta$ has a definite sign,
precisely that of $-k$. Therefore, if we start {}from a positive $\eta(0)$,
and $k>0$, $\ddot\eta(\tau)$ will be negative for as long as $\eta(\tau)>0$.
Then, if $\dot\eta(0)<0$, $\dot\eta(\tau)$ will grow in absolute value
for as long as $\eta(\tau)>0$. In conclusion, $\eta$ will vanish for some
finite value of $\tau=\tau_0$. On the other hand, if $\dot\eta(0)>0$
(always for $k>0$), $\eta(\tau)$ need not vanish. The numerical 
analysis supports these conclusions.

In fig. \figkdneg\ we plot $\eta(\tau)$ and $f(\tau)$ for $k=2$ with initial
conditions such that $\eta(0)\dot\eta(0)<0$. We observe that the
behaviour \singu\ is reproduced with reasonable accuracy.
In fig. \figkcneg\ we again consider $\eta(0)\dot\eta(0)<0$,
but this time for $k=4$.

Fig. \figkdpos\ depicts $\eta(\tau)$ and $f(\tau)$ for $k=2$ with
initial conditions such that $\eta(0)\dot\eta(0)>0$. We then 
find that the asymptotic regime \solasim\ is very quickly 
reached. Similarly for $k=4$ (fig. \figkcpos).

In fig. \figkdtot\ we portray a solution for $k=2$ which starts {}from
nothing at the big crunch $\eta=\tau=0$ and grows steadily 
up to an oscillatory solution, which is to be understood as
a graviton or dilaton created ex nihilo.

Next figure, fig. \figktra, presents a numerical solution for $k=-3/2$,
close to a vanishing point of $\eta$,
in order to show that the analysis is general. It is clear that it
obeys equation \exun\ (modulo an unimportant scaling factor). 
We proceed to show in fig. \figktrb\ a numerical solution, also for
$k=-3/2$, but this time illustrating the asymptotic regime \solasim.

There are other possible modes of the string, and in fact
our whole analysis easily generalizes to higher winding 
modes: take the following ansatz,
$$\eqalign{
\eta&=\eta(\tau)\,,\cr
X^1&=f_n(\tau)\cos n\sigma\,,\cr
X^2&=f_n(\tau)\sin n\sigma\,,\cr
X^i&={\rm cons.}\,,\quad i\geq3\,.\cr}\eqno\eq$$
Apart {}from the exponential solutions, which take the form
$$\eta(\tau)=e^{\pm n\tau/\sqrt{-k-1}}\,,\quad 
f(\tau)={1\over{{\sqrt{-k}}}}e^{\pm n\tau/\sqrt{-k-1}}
\,,\eqno\eq
$$
only the $\tau\to\infty$ asymptotic behaviour is slightly
modified in the context of this generalized ansatz. We find
that the analogue of \solasim\ is
$$\eqalign{
\eta(\tau)&\buildrel{{\tau\to+\infty}}\over\sim \tau^{2/(2+k)}\,,\cr
f(\tau)&\buildrel{{\tau\to+\infty}}\over\sim 
{2\over{n(2+k)}} \tau^{-k/(2+k)} 
\cos(n\tau+\varphi)\,.\cr}\eqn\solasn$$

\chapter{Strings propagating in stringy cosmological spacetimes}

There are a number of interesting string vacua of this
kind to be found in the literature [see (\tseyt) for a review].
These geometries can be either considered in the 'string frame' or in
the 'Einstein frame'. The physics {\it changes} with the frame.
We choose to work in the Einstein frame. The reason being that in such
frame the spacetime metric in the effective field theory action is the
{\bf same} metric that appears in the string action for test strings [\SV].
In the string frame, the metric in the effective action contains the
dilaton field whereas the dilaton never couples with classical test strings. 

We shall assume $D = 4$ uncompactified dimensions for this stringy
universe as it was before the case for the FRW geometries (sec.2).
In addition, we assume some extra compactified degrees of freedom
contributing with an amount $c < 0$ to the central charge. No further
assumptions will be made. In fact the value of $|c|$ gets absorbed
in the choice of unit of length. 

We consider in particular cosmological backgrounds in maximally 
symmetric, conformally flat space, that is to say, with metric of 
the form \metric. Making the same ansatz as before \ansatz, we again have 
equations \deru\ as equations of motion and constraint.
Notice that the invariant string size  is given by
$$S(\tau)=f(\tau)\; R(X^0(\tau))\,.\eqno\eq$$

Let us first examine the linear dilaton, flat string frame
metric solution of Myers [\myers], in Einstein's frame [\SV, \tseyt],
(with the correct constant rescaling of the spatial components):
$$\eqalign{
{\rm d}s_E^2 &=({\rm d}X^0)^2-(X^0)^2\sum_{i=1}^{D-1} ({\rm d}X^i)^2=\cr
&=e^{2\eta}\left[{\rm d}\eta^2-\sum_{i=1}^{D-1} ({\rm d}X^i)^2\right]
\,,\cr}
\eqno\eq$$
whence the equations of motion and constraint
$$\eqalign{
\ddot f + 2 \dot\eta\dot f + f &= 0\,,\cr
\ddot \eta + 2 \dot f^2 &=0\,,\cr
\dot\eta^2-\dot f^2 - f^2 &=0\,,\cr}\eqn\myereq$$
for ring-like solutions, following the by now usual ansatz.
It has to be observed that $\eta$ does not enter these equations save
through its derivatives. We thus see that $\eta$ can be shifted by
an arbitrary constant. Disregarding this
constant, we are actually restricted to a single non-linear
second order ordinary differential equation for $f$, up to 
the choice of a sign for $\dot\eta$. There is no scaling
freedom in this case.

In an analogous manner to the analysis carried out before, we
obtain that the asymptotic behaviour of the solutions for $\tau\to+\infty$
is of the form
$$\eqalign{
f(\tau) &\buildrel{{\tau\to+\infty}}\over={1\over\tau}
\cos(\tau+\varphi)+O(1/\tau^2)\,,\cr
\eta(\tau) &\buildrel{{\tau\to+\infty}}\over= \eta_0 + \ln\tau +
O(1/\tau)\,.\cr}\eqn\asymyers
$$
Here, the oscillation amplitude for $f(\tau)$ is fixed for all
ring solutions. Since the asymptotic string energy is proportional to
 $e^{ \eta_0}$, it is then this arbitrary parameter which effectively plays
the r\^ole of asymptotic string amplitude.

The string size  is 
$$S(\tau) = e^{\eta(\tau)} f(\tau)\,,\eqno\eq$$
and asymptotically for  $\tau\to+\infty$
$$S(\tau)\buildrel{{\tau\to+\infty}}\over\sim 
e^{\eta_0}|\cos(\tau+\varphi)|\sim {1\over \sqrt2}e^{\eta_0}
\,.\eqno\eq$$
The string is of a bounded size, whereas the universe is expanding
in this regime.

Observe that a) $\ddot\eta$
is always negative; b) $\dot\eta=0$ (which implies
$f=0$, $\dot f=0$) is a critical point of this 
two dimensional dynamical system. This means that $\dot\eta>0$ and
$\dot\eta<0$ are disconnected regions of configuration space, as was pointed
out before. On the other hand, time reversal invariance means that
they can be mapped into each other, so we concentrate on $\dot\eta>0$.
Because of a), $\dot\eta$ will tend towards zero as $\tau$ grows, 
which corresponds to \asymyers. Backwards in time, though, it will
grow indefinitely. This growth might be up to some finite $\tau_0$,
in which case, there will be a divergence
at this point, of the form
$$\eqalign{
f(\tau) &\buildrel{\tau\to\tau_0}\over\sim{1\over2}\ln(\tau-\tau_0)\,,\cr
\eta(\tau) &\buildrel{\tau\to\tau_0}\over
\sim \eta_0 +{1\over2} \ln(\tau-\tau_0)\,.\cr}\eqn\singumyers$$
The string collapses in this case, but more slowly than the
spacetime: 
$$S(\tau)\buildrel{\tau\to\tau_0}\over
\sim{\sqrt{\tau-\tau_0}\over 2}|\ln(\tau-\tau_0)|\,.\eqno\eq$$

Both types of asymptotic behaviours, \asymyers\ and \singumyers, are
observed in the numerical solution depicted in figure \figmyers.

The string energy \eneg\ tends to $ {1 \o {\a '}}\; e^{ \eta_0} $
for $X^0 \to \infty $ and it is redshifted as 
$${{e^{2 \eta_0}} \o { 2 \a' R(X^0)}} $$
 near the big bang singularity as in eq.\ensin .

As a conclusion, the behaviour of ring-like string solutions in
this spacetime is analogous to that found for FRW spacetimes,
by making $k$ tend formally towards $+\infty$.
An important difference between this spacetime and the FRW solutions
of general relativity lies in the fact that the ``big crunch'' singularity
for the case at hand is a null singularity [\horo], whereas it is 
spacelike for FRW spacetimes. 
The collapse of the strings into these singularities is also of
different kinds: while $S(\tau)/R(\tau)$ tends powerlike to zero  for
FRW spacetimes $(k>0)$, it is logarithmically divergent for the
stringy solution under consideration.

Another interesting spacetime is the isotropic case of 
the one obtained by Mueller [\mueller], which
is asymptotically identical to the previous one for $X^0\to+\infty$, but
which posseses simple polynomial curvature singularities for two values of
$X^0$. In Einstein's frame, it reads
$$\eqalign{
{\rm d}s_E^2 &=({\rm d}X^0)^2- R(X^0,a)^2\sum_{i=1}^{D-1} ({\rm d}X^i)^2\cr
&=({\rm d}X^0)^2-R_0^2 ~(X^0-a)^{(\sqrt3 +1)/\sqrt3}
(X^0+a)^{(\sqrt3 -1)/\sqrt3}\sum_{i=1}^{D-1} ({\rm d}X^i)^2 ,\cr}
\eqn\metm
$$
for $D=4$. Here $c < 0$ as explained above and $a$ is an arbitrary constant.
Observe that
$$R(\lambda X^0,a)=\lambda R(X^0,a/\lambda)\,,$$
{}from which we can fix $a=1$ without loss of generality,
and $R_0$ can also be set to 1 by an adequate constant rescaling
of the spatial coordinates. Making again the ring-like
ansatz \ansatz, the equations of motion \deru\ are 
consequently :
$$\eqalign{
\ddot f + & 
2\left({{c_1}\over{X^0-1}}+ {{c_2}\over{X^0+1}}\right)\dot X^0\dot f
+ f=0\,,\cr
\ddot X^0 + & (X^0-1)^{2c_1} (X^0+1)^{2c_2}\left({{c_1}\over{X^0-1}}+ 
{{c_2}\over{X^0+1}}\right)\left(\dot f^2 - f^2\right)=0\,,\cr
(\dot X^0)^2 - & (X^0-1)^{2c_1} (X^0+1)^{2c_2}\left(\dot f^2 + f^2\right)
=0\,.\cr}\eqn\eqmuller$$

Here,
$$
c_1={1\over2}(1+{1\over{\sqrt3}})~,~ 
c_2={1\over2}(1-{1\over{\sqrt3}}) \eqn\cucd
$$
In $D$ space-time dimensions one just has to replace $\sqrt3$ in 
eqs.\metm -\cucd\  by $\sqrt{D-1}$.

{}From the expression of the metric one can conclude that in
the asymptotic region $X^0\to+\infty$ the behaviour is analogous to
that of Myers' spacetime. That is to say,
$$\eqalign{
X^0 &\buildrel{{\tau\to+\infty}}\over\sim p\tau\,,\cr
f &\buildrel{{\tau\to+\infty}}\over\sim 
{1\over{\tau}}\cos(\tau+\varphi)\,.\cr}\eqn\myasi
$$
where $p$ is an arbitrary positive parameter proportional
to $e^{\eta_0}$.
On the other hand, it is clear that $X^0=\pm1$ are singularities
of the metric.
The curvature scalar tends to $\infty$ as those points are approached,
and the metric is imaginary for $-1<X^0<1$.
It is also evident that $X^0=\pm1$ are critical planes for the
set of ordinary 
differential equations \eqmuller.
Equations \eqmuller\ are
autonomous, so there is invariance under time translation, and
we can choose  $\tau = \tau_0$ as the point for which $X^0=\pm1$.
 In such case, the
following asympotic behaviour is obtained by demanding that the
solutions be regular for $X^0\to1^+$ when $\tau \to \tau_0$ (we choose
for simplicity $\tau_0 = 0$) :
$$\eqalign{
X^0(\tau) & \buildrel{\tau\to 0}\over\sim 1+b_1\tau^{(1/c_2)}\,,\cr
f(\tau) & \buildrel{\tau\to 0}\over\sim 
\left({{b_1}\over2}\right)^{c_2} {1\over{c_2}}
\left[1+O(\tau^2)\right]\,.\cr}\eqno\eq$$
This leads to a string size squared that tends to zero at the 
same rate as the spacetime.

Allowing for singular solutions, we obtain for 
$X^0(\tau)\to1^+$ as  $\tau \to 0$
$$\eqalign{
X^0(\tau) & \buildrel{\tau\to 0}\over\sim 1+b_1~\tau^{1/(c_1+1)}\,,\cr
f(\tau) & \buildrel{\tau\to 0}\over\sim 
\left({{b_1}\over2}\right)^{c_2} {1\over{c_2}}~
\tau^{c_2/(c_1+1)}~\left[1+O(\tau^2)\right]\,.\cr}\eqn\singmueller$$
in which case the string collapses as $\tau\to 0^+$ ($b>0$)
with 
$${{S(\tau)}\over{R(\tau)}}\buildrel{\tau\to 0}\over\sim 
\tau^{c_2/(c_1+1)}\,.\eqno\eq$$
Alternatively, and when consider $X^0=1$ as the big bang (and
$X^0=-1$ as the big crunch), the string is coming out of the
big bang faster than the expansion rate of the universe.

Passing to $X^0\to-1$, we see that there are solutions of
the same type as above, given by the exchange of $c_1$ with $c_2$,
and such that $X^0\to-1^-$.

 To illustrate these results, we portray in fig.\figmuellera\ a numerical
solution in which the behaviour for $\tau\to+\infty$ is particularly 
clear, and next in fig.\figmuellerb\ a closer look at the limit 
$X^0\to1^+$, for a
different numerical solution. It is seen that the singular behaviour
\singmueller\ is obeyed.

\chapter{Strings propagating in black hole spacetimes}
\def\rs{r_\sigma}
\def\rt{r_\tau}
\def\rss{r_{\sigma\sigma}}
\def\rtt{r_{\tau\tau}}
\def\ts{t_\sigma}
\def\ta{t_\tau}
\def\tss{t_{\sigma\sigma}}
\def\taa{t_{\tau\tau}}
\def\ps{\phi_\sigma}
\def\pa{\phi_\tau}
\def\pss{\phi_{\sigma\sigma}}
\def\paa{\phi_{\tau\tau}}
\def\tes{\theta_\sigma}
\def\tea{\theta_\tau}
\def\tess{\theta_{\sigma\sigma}}
\def\teaa{\theta_{\tau\tau}}

The motion of
classical test strings in a black hole background metric is also particularly
interesting. Previous works on classical strings in black hole
backgrounds considered infinite strings [\lm ], static solutions 
[\fal ], charged strings [\lar ] and perturbations on static solutions
[\fl  ].

We start with the Schwarzschild metric
$${\rm d}s^2=R_s^2\left[-\left(1-{1\over{r}}\right){\rm d}t^2 + {{{\rm d}r^2}\over{\left(1-{1\over{r}}\right)}} + r^2({\rm d}\theta^2+\sin^2\theta\>{\rm d}\phi^2)\right]\,,\eqno\eq$$
where $R_s=2m$ is the Schwarzschild radius, and the radial and ``temporal''
coordinates are $R=rR_s$ and $T=tR_s$.
The equations of motion of a classical test string in such a 
background, and in the conformal gauge, are
$$\eqalign{
\rs\ts - \rt\ta + r(r-1)(\tss-\taa)&=0\,,\cr
r\sin\theta\>(\paa-\pss) + 2 r\cos\theta\>(\pa\tea-\ps\tes) + 2 
(\rt\pa-&\rs\ps)=0\,,\cr
2r(\teaa-\tess)+ 4(\rt\tea-\rs\tes) - r\sin2\theta\>(\pa^2-\ps^2)&=0\,,\cr
{{2r}\over{1-r}}(\rtt-\rss) - {1\over{r^2}}(\ta^2-\ts^2) + 2r ( \tea^2 -\tes^2)+\quad&\cr
+2r\sin^2\theta\>(\pa^2-\ps^2) &+ {1\over{(r-1)^2}}(\rt^2-\rs^2)=0\,.\cr
}\eqn\eqgen$$
The constraints are
$$\eqalign{
{{1-r}\over{r}}(\ts^2+\ta^2) + {r\over{r-1}}(\rt^2+\rs^2)
+& r^2 (\tea^2+\tes^2) + r^2\sin^2\theta\>(\pa^2+\ps^2)=0\,,\cr
{{1-r}\over{r}}\ta\ts +  {r\over{r-1}} \rt\rs + & r^2\tea\tes 
+ r^2\sin^2\theta\>\pa\ps=0\,.\cr}
\eqn\eqconst$$
\subsection{Rotationally symmetric, static 2+1 spacetime}

Consider for a while constant $\theta$ in the equations above. This case is
similar to the motion of a string in a conformally flat, 
rotationally symmetric, static 2+1 spacetime,
$${\rm d}\tilde s^2=A^2(\rho)\left(-{\rm d}t^2 + {\rm d}\rho^2+ 
\rho^2{\rm d}\phi^2\right)\,.\eqno\eq$$
If we assume, as is natural, that $\phi=n\sigma$, then either
$A(\rho)\propto 1/\rho$, or $\rho=\rho(\tau)$. In the second
case, $\rho=\rho(\tau)$, we have that either $t=t(\tau)$ or
$t=t(\sigma)$. Both these cases can be solved by quadratures.
The less physical one (which would correspond to compactified
$t$ coordinate), $t=t(\sigma)$, produces the solution
$$\eqalign{
\rho&=(\gamma/n) \cos(n\tau+\nu)\,,\cr
t&=\gamma\sigma+ \delta\,,\cr}\eqn\solunph$$
with $\gamma,\nu$ and $\delta$ arbitrary constants.

The more interesting case $\rho=\rho(\tau)$, $t=t(\tau)$,
has the following solution ($\omega=\log A$):
$$\eqalign{
\int^\rho {{{\rm d}\rho}\over{\left(c^2 e^{-4\omega} - 
n^2 \rho^2\right)^{1/2}}}
&=\pm\tau + d\,,\cr
t&=c\int e^{-2\omega}{\rm d}\tau + g\,,\cr}\eqn\solph$$
again with arbitrary constants $c,d,g$.

Our problem of interest, in a Schwarzschild background, cannot be
put in this form. However, now that the analysis has been carried
out for the conformally flat case, it is straightforward to generalize
it to applicable expressions.

Consider then the following metric:
$${\rm d}\tilde s^2=A^2(\rho)\left(-{\rm d}t^2 + {\rm d}\rho^2+ 
b(\rho)^2\rho^2{\rm d}\phi^2\right)\,.\eqno\eq$$
If we assume, as before, that $\phi=n\sigma$, we have that 
either $Ab\propto1/\rho$, or $\rho=\rho(\tau)$, in which 
event, either $t=t(\sigma)$ or $t=t(\tau)$, as before, and the
solutions are also given by quadratures: either
$$\eqalign{
\int^\rho{{{\rm d}\rho}\over{\left(\gamma^2-n^2 b^2 \rho^2\right)^{1/2}}}
&=\pm\tau+\nu,,\cr
t&=\gamma\sigma+ \delta\,,\cr}\eqn\solunphd$$
with $\gamma,\nu$ and $\delta$ arbitrary constants, or
$$\eqalign{
\int^\rho {{{\rm d}\rho}\over{\left(c^2 e^{-4\omega} - 
n^2 b^2\rho^2\right)^{1/2}}}
&=\pm\tau + d\,,\cr
t&=c\int e^{-2\omega}{\rm d}\tau + g\,,\cr}\eqn\solphd$$
again with arbitrary constants $c,d,g$.

These solutions can now be applied to the case of equatorial
motion of a string in a Schwarzschild black hole background.
As a matter of fact, assume just that $\theta$ is constant.
This implies that either $\theta=\pi/2$ (since $\theta$ is restricted
between 0 and $\pi$), or $\phi_\tau^2=\phi_\sigma^2$. Taking the
axisymmetric solutions ($\phi=n\sigma, n =$integer), it is clear that only
the equatorial motion makes sense, as is also physically evident.
Let us concentrate on such an equatorial, axisymmetric motion.
Applying formula \solphd, and the restriction
that $r$ cannot be smaller than 0, we are led to 
$$\eqalign{
t&=t_0={\rm cons.}\,,\cr
r&=\cos^2\left({{n\tau + \delta}\over2}\right)\,.\cr}\eqn\equin$$
This equation describes a string propagating inside the Schwarzschild
horizon $r=1$. 
Let us remember that $r$ takes on a time-like character inside the 
horizon, whereas $t$ is space-like.
It is thus better to study this solution in Kruskal-Szekers (KS) coordinates
[\blackb]:
$$\eqalign{
u&=\sqrt{1-r}~ e^{r/2}~ \sinh(t/2)\,,\cr
v&=\sqrt{1-r}~ e^{r/2}~ \cosh(t/2)\,.\cr}
\eqn\krus$$
The metric takes in these coordinates the following form:
$${\rm d}s^2={4\over{r}}R_s^2~e^{-r}\left(-dv^2 + du^2\right)
+R_s^2~ r^2~({\rm d}\theta^2+\sin^2\theta\>{\rm d}\phi^2)\,.\eqno\eq$$
The coordinate $v$ is a time-like coordinate, and $u$ spacelike.
We find {}from equations \equin-\krus:
$$\eqalign{
r&=\cos^2\left({{n\tau + \delta}\over2}\right)\,,\cr
u&=\tanh(t_0/2)~ v\,.\cr}
\eqn\solkrus$$
That is, the string falls into the singularity $r=0$, with 
constant speed $\tanh(t_0/4)<1$ with respect to KS coordinates
(see fig. \figbhint).
Solution \solkrus\ (for $n=1$ and $ \delta = 0$) 
starts at, say, $\tau=0$, on the horizon, and
bounces back to $r>0$ after $r$ has vanished for $\tau=\pi$.
This behaviour may be  interpreted as follows:
The motion outwards the singularity is unphysical.
Which is to say that the string motion ends when the singularity
$r=0$ is reached at $\tau=\pi$.
Moreover, the invariant string size vanishes as $ (\tau - \pi)^4 $
when the singularity is approched.

The other (unphysical) solution \solunphd, leads to 
$$\eqalign{
t&=\gamma\sigma + \delta\,,\cr
\int^r{{r{\rm d}r}\over{\left(r-1\right)^{1/2}
\left(\gamma^2(r-1)- n^2 r^3\right)^{1/2}}}&=\pm\tau + \beta\,,\cr}
\eqno\eq$$
which is not so interesting given the periodicity we impose on
$\sigma$.
\subsection{Axisymmetric ansatz}

The Schwarzschild manifold not being a symmetric space,
the string equations of motion and of constraint, \eqgen\ and \eqconst,
are not integrable there. In order to separate the equations into
ordinary differential equations, we make a simple ansatz
with a symmetry compatible with the time evolution. Our
concrete ansatz is to choose an axisymmetric (ring) configuration
as follows:
$$\phi=n\sigma\,,\quad\theta=\theta(\tau)\,,\quad t=t(\tau)\,,\quad
r=r(\tau)\,.\eqn\anbh$$
where $n =$ integer. This ansatz inserted in equations \eqgen\ 
and \eqconst\ produces the 
following set of equations:
$$\eqalign{
\ddot r - (r-3/2)\dot\theta^2 + n^2\sin^2\theta (r-1/2)&=0\,,\cr
r\ddot\theta + 2\dot r\dot\theta +n^2 r\sin\theta \cos\theta&=0\,.\cr}
\eqn\anbeq
$$
with a conserved quantity
$$e^2=\dot r^2+r(r-1)(\dot\theta^2+n^2\sin^2\theta)\,,\eqn\anbco$$
and $t$ given by
$$\dot t={{er}\over{r-1}}\,.\eqn\anbtd$$
Note that these equations are invariant under the
change $\tau\to-\tau$, $t\to-t$.
Since $\dot t > 0$ outside the horizon, $t(\tau)$ is a 
monotonous function, and we can use
either $\tau $ or $t$ to study the time evolution for $r > 1$.

The string energy  is found to be in this case, 
$$E(t)\equiv -P_0 = -{{G_{00}}\o{\a'}}{{dX^0}\o{d\tau}} =
{{R_s\;e}\over{\a'}}\,.\eqn\enan$$
where we used eq.\tens :
 $$P^0 = \int d^{D-1}X \sqrt{-G}~ T^{00}(X)$$
and $\a'$ is the string tension.

The invariant length of the string in this case is 
$${\rm d}s^2=R_s^2n^2r^2\sin^2\theta\>(-{\rm d}\tau^2+{\rm
d}\sigma^2)\,.
\eqn\taneg
$$
A useful equation, satisfied by solutions of these 
equations, is
$$\left({{{\rm d}^2}\over{{\rm d}\tau^2}}+ n^2\right)(r\sin\theta)=
{1\over2}(n^2\sin^2\theta - 3\dot\theta^2)\sin\theta\,.\eqn\osceq$$

Let us now examine the possible asymptotic behaviours of these
equations in different regimes. A first interesting question
is the existence of collapsing solutions and the corresponding
critical exponents.
On computation, we find two possible collapsing behaviours {}from
\anbeq-\anbco, with the adequate choice of origin for $\tau$
for  $\tau \to 0$:
$$\eqalign{
r&\buildrel{{\tau\to 0}}\over\sim\alpha \; \tau^{2/5}\,,\cr
\theta&\buildrel{{\tau\to 0}}\over\sim
\theta_f+ 2\; \sqrt{\alpha} \; \tau^{1/5}\,,\cr}
\eqn\caidau
$$
where $\a $ is a constant and
$$\eqalign{
r\buildrel{{\tau\to 0}}\over\sim &~e\tau +
 {{n^2 \sin^2\theta_f}\over4}~\tau^2 - 
{{n^2 e\sin^2\theta_f}\over6}~\tau^3 + O(\tau^4)\,,\cr
\theta\buildrel{{\tau\to 0}}\over\sim &~\theta_f - 
{{n^2 \sin(2\theta_f)}\over{12}}~\tau^2 
+{{n^2\sin^3\theta_f\cos\theta_f}\over{726}}~\tau^3+ O(\tau^4)\,.\cr}
\eqn\caidad$$
This last one is obviously subdominant with respect to eq. \caidau .

Consider now the regime given by large $r$ and $\dot\theta^2/r$ small.
{}From equations \anbco\ and \osceq, we have that, for $|\tau|\to+\infty$,
$$\eqalign{
r&\buildrel{{\tau\to +\infty}}\over\sim p|\tau|\,,\cr
\theta&\buildrel{{\tau\to +\infty}}\over\sim \theta_0 - 
{{m}\over{p}}{{\cos(n\tau+\varphi_0)}\over\tau}\,,\cr}
\eqn\bhinfas$$
together with
$$e^2=p^2+n^2 m^2\,,$$
and $t\buildrel{{\tau\to +\infty}}\over\sim e\tau$. 
Here $\theta_0$ is such that $\sin\theta_0=0$,
i.e., $\theta=l\pi$ with $l$ an integer.
We could here understand $p$ as an asymptotic radial momentum, $e$
the energy, and
$n m$ the mass of the string. The latter is determined by the amplitude 
of the string oscillations. 

We find for large $\tau$ that
$$\eqalign{
x=r\sin\theta \cos\phi& = (-1)^{l+1} ~m~ \cos(n\tau+\varphi_0)\cos n\sigma\,,\cr
y=r\sin\theta \sin\phi& = (-1)^{l+1} ~m ~\cos(n\tau+\varphi_0)\sin n\sigma\,.\cr}
\eqn\coorxy$$
In this region, spacetime is minkowskian, and we can recognize \coorxy\ as 
the $n^{{\rm th}}$ excitation mode of a closed string. For $|n|=1$ this
corresponds at the quantum level to a graviton and/or a dilaton.
Notice that $m$ is  the amplitude of the string oscillations.

The string size is here $ S(\tau) = R_s r(\tau) ~ |\sin\theta(\tau)|
$. We find from eqs.\caidau -\bhinfas\ that 
$$\eqalign{
S(\tau)\buildrel{{\tau\to +\infty}}\over\sim& R_s m |\cos(\tau +
\varphi)|\buildrel{{\tau\to +\infty}}\over\sim {{R_s m}\o {\sqrt2}}\cr
S(\tau)\buildrel{{\tau\to +0}}\over\sim&R_s \a
\sin\theta_f~\tau^{2/5}. \cr}
\eqn\tamul
$$

 Whenever the string is not swallowed by the black hole, equation
\bhinfas\ describes both the incoming and outgoing regions
$\tau\to\pm\infty$. However, the mass $m$, the momentum $p$ and 
the phase $\varphi_0$ are in general different in the two
asymptotic regions. This is an illustration of a rather
general phenomenon noticed at the quantum level: particle 
transmutation [\tras ]. This means that the excitation
state of a string changes in general when it is scattered
by an external field  like a black hole. Within our
classical ansatz \anbh, the only possible changes are in
amplitude (mass), momentum, and phase.
It can be seen numerically that the excitation state is 
indeed modified
by the interaction with the black hole. 

Due to the structure of our ansatz, the string, if it is not
absorbed for some finite $\tau$, may return to $z=+\infty$ 
($\theta_f=0\pmod{2\pi}$), where it started at $\tau=-\infty$,
or go past the black hole towards $z=-\infty$ ($\theta_f=\pi\pmod{2\pi}$)

An special case of interest is that of solutions such that
$r(\tau)=r(-\tau)$. It follows that 
$\dot\theta^2(\tau)=\dot\theta^2(-\tau)$, from which
$\theta(\tau)=\Delta-\epsilon\theta(-\tau)$,
with $\epsilon$ a sign. We 
then see that $\Delta$ is restricted to multiples of 
$\pi$ if the string does not fall into the black hole. 
If it is an odd multiple, we understand that 
the string has circled round the black hole a number 
of times and then  has continued to infinity, whereas
when it is an even multiple, the string bounces back 
after some dithering around the black hole. This analysis can
be extended to all solutions.

Let us now analyze the absorption of the string by the 
black hole. If $r$ starts at $+\infty$ for $\tau=-\infty$ and
decreases ($\dot r<0$), $\dot r$ must change sign at the 
periastron at time $\tau_0$. Otherwise the singularity at
$r=0$ will be reached. Furthermore, $\ddot r(\tau_0)>0$.
We see {}from the first equation in the set \anbeq\ that 
this implies that $r(\tau_0)>3/2$. In other words, if the
string  penetrates the $r<3/2$ region, it will necessarily
fall into the singularity. In yet another paraphrase,
there is an effective horizon
for ring string solutions. The surface $r=3/2$ is necessarily
contained within this horizon. Let us recall that for massless
geodesics the  effective horizon is a sphere of radius 
$r = {3 \over 2} \sqrt3 $.

To illustrate these points, we adjunct some figures.
They depict the motion of ring-like strings described by
equations \anbh\ through \anbtd. We numerically integrate
equations \anbeq\ {}from large negative $\tau$, where the 
asymptotic behaviour \bhinfas\ holds. We choose $\theta_0$,
$n=1$, and vary the values of $p,m$ and $\varphi_0$. Depending on
this last set of three values  the string is absorbed or not
by the black hole.

In fig. \figbhfal\ we show an example of direct fall (i.e.,
with no bobbing around the black hole). 
In order to
compare with this one, we next portray (fig. \figbhpas) a case where the string
goes past the black hole before returning to it and collapsing. The
clearest view of this event is given by fig. \figbhpas d, which depicts
$z=r\cos\theta$ as a function of  $\rho=r\sin\theta$.

Fig. \figbhreb\ is 
dedicated to a non-falling string. It is particularly
interesting to point out that the excitation state has been
changed by scattering by the black hole, as can be clearly seen {}from
the third graph in this figure, which depicts $r\sin\theta$ as a function
of $\tau$. 
We see that the oscillation amplitude is larger after the collision
than before. This means that the outgoing string mass is larger than the
ingoing string mass. Hence, particle transmutation in the sense
of ref.[\tras ] takes place here.

In the fourth of this series, fig. \figbhreb d,
we portray $z=r\cos\theta$ against $\rho=r\sin\theta$. It is to be remarked
that the string bounces (the lower end of the picture), then oscillates
around the black hole, and finally escapes to infinity.

That the string be absorbed or not by the black hole is dictated by
whether it comes or not within the effective horizon,
as mentioned above. This, in turn, is crucially dependent on the phase
$\varphi_0$ chosen as part of the initial data ($\tau\to-\infty$).
Whatever value the mass (amplitude) $m$ and the momentum $p$ take,
there is always some interval of values of $\varphi_0$ for which
the string will be absorbed by the black hole.

Besides numerical experiments, this behaviour follows {}from the simple
fact that a change of the initial phase $\varphi_0$ would displace the 
string worldsheet, thus possibly bringing it closer to the black hole.

\chapter{Strings propagating in stringy black holes}
Among the tree level string vacua of this type we shall only consider
non-rotating, $3+1$-dimensional black hole spacetimes of the type 
first presented in [\garymae].

Before coming to those, though, let us write the equations of motion
corresponding to a ring configuration of a classical string in a metric
of the form (in Einstein's frame)
$${\rm d}s_E^2=-A(r) {\rm d}t^2 + {1\over{B(r)}}{\rm d}r^2 + 
C(r)({\rm d}\theta^2 + \sin^2\theta {\rm d}\phi^2)\,.\eqn\metgenan$$
Our ansatz \anbh\ can be used for this general metric. The
equations of motion and constraint are as follows:
$$\eqalign{
\ddot r =& {1\over2}\left\{{{AB'-BA'}\over{AB}}~\dot r^2 
+ BC\left({{C'}\over{C}}-{{A'}\over{A}}\right)\dot\theta^2-
n^2BC\sin^2\theta\left({{C'}\over{C}}+{{A'}\over{A}}\right)\right\}\,,\cr
\ddot\theta =& -{{C'}\over{C}}~\dot r~\dot\theta-n^2\sin\theta\cos\theta\,,\cr
\dot t =& {{e}\over{A}}\,,\cr
e^2 =& {{A}\over{B}}~\dot r^2 + AC~\dot\theta^2 + AC n^2 \sin^2\theta\,.\cr}
\eqno\eq$$
Here $e$ is a constant of motion and the primes denote derivatives 
with respect to $r$ for $A,B$ and $C$, which are functions of $r$.
The string size squared is in this case
$$S(\tau)^2=C(r)\;n^2\sin^2\theta\,.\eqno\eq$$
As in sec. 4 [eq.\enan ], the string energy is given here by 
$E(t)={{e}\over{\a'}}$

More specifically, let us consider the following set of metrics [\garymae]:
$$\eqalign{
A(r)=B(r) = & ~{{(r-r_+)(r-r_-)}\over{r^2-r^2_0}}\,,\cr
C(r)= & ~r^2 - r^2_0\,.\cr}\eqno\eq$$
This corresponds to the presence of an electric and
a magnetic charge, to which the string does not couple save
for their effect on the metric. The parameters of the 
metric are derived from the mass $M$ and the charges $Q_E$
(electric) and $Q_M$ (magnetic) as follows.
$$r_0={{Q_M^2-Q_E^2}\over{2M}}\,,\quad 
r_\pm=M\pm(M^2+r_0^2-Q_E^2-Q_M^2)^{1/2}\,.\eqno\eq$$
If $Q_E^2>Q_M^2$, the string coupling constant goes to
zero close to the singularity $r=|r_0|$, and we would
thus be justified in considering this a good approximation to an
exact string vacuum. Remember that all these metrics are solutions to
the tree-level effective action.

These metrics are globally very similar to Reissner-Nordstr\oo m's (RN)
when $Q_E$ and $Q_M$ are both different from zero. There are
two horizons: an event horizon at $r_+$ and an inner horizon
at $r_-$; and a singularity at $r=|r_0|$. As a matter of 
fact, RN appears as the special case $Q_E=Q_M$. 

A very interesting phenomenon appears for these spacetimes 
for the equatorial motion
of a ringlike string. We obtain the following exact solution to the 
equations of motion:
$$\eqalign{
\phi=  n\sigma\,,\quad\theta=\pi/2\,,&\quad\dot t=e/A(r(\tau))\,,\cr
r(\tau)= M + & a \cos(n\tau+\varphi)\,,\cr}\eqn\eqsol$$
with the following constraints on $a$, coming from the positivity
of $e^2 = a^2 + Q_E^2+Q_M^2- M^2- r_0^2 $,
 and from imposing that $r(\tau)\geq|r_0|$:
$$M-|r_0|\geq|a|\geq(M^2+r_0^2-Q_E^2-Q_M^2)^{1/2}\,.$$
{F}rom this constraints it follows that 
a) the periastron lies
{\it inside} the inner horizon: $r_{\rm min}\leq r_-$; b)
the apoastron lies {\it outside} the event horizon: $r_{\rm max}\geq r_+$.
c) these last two formulae are only saturated when 
$a=(M^2+r_0^2-Q_E^2-Q_M^2)^{1/2}$.
We are thus presented with the situation that a classical
string enters the nonstatic region, goes on to face the singularity, and
without falling into it bounces back out again. This looks clearly
analogous to a timelike curve that
crosses from one asymptotically flat region to another, of
those present in 
the maximal analytic extension of this spacetime. In other words,
the string travels to other universes through the wormholes, and
continues to do so indefinitely. The extreme case
$a=(M^2+r_0^2-Q_E^2-Q_M^2)^{1/2}$ corresponds to $e^2=0$,
{from} which it follows that in this case the constant $t$ trayectories
within the region $r_-\leq r\leq r_+$
are the ones the string takes, never entering neither the
asymptotically flat regions, nor the regions connected with the
singularity. The other extreme case, $a=M-|r_0|$, corresponds
to a string that reaches the singularity.

Let us now examine the asymptotic behaviour of ringlike configurations
of classical strings far away from the black hole and very close to 
it. Far away from it, the metric is Minkowski's up to $1/r$ terms.
Therefore the asympotic behaviour of ring solutions is the same as
for Schwarzschild's background, formula \bhinfas. More interesting
is the possibility of collapse onto the singularity.
Whenever $r_+$ and $r_-$ are real, as we are
supposing all along, it will always be true that
$(|r_0|-r_+)(|r_0|-r_-)\geq0$. This inequality will only be 
saturated when $Q_E=Q_M=0$, which would place us in the already
analyzed Schwarzschild case.
Therefore, consider $(|r_0|-r_+)(|r_0|-r_-)>0$. 
We find that the string generically cannot reach
the singularity $r=|r_0|$. Assume that it does reachit, say, for $\tau
\to 0$ .
It follows from the equations of motion,
$$\eqalign{
\ddot r=&\left[{2\o{r}}(r-r_+)(r-r_-)-r +
{1\o2}(r_++r_-)\right]\dot\theta^2
- \left[r-{1\o2}(r_++r_-)\right]n^2\sin^2\theta\,,\cr
\ddot\theta=& -{{2r}\o{r^2-r_0^2}}\;\dot r~\dot\theta -
 n^2\sin\theta\cos\theta\,, \cr}\eqno\eq$$
that the generic behaviour is 
$$\eqalign{
r\buildrel{\tau\to 0}\over\sim&\;|r_0| + c \tau^\a\,,\cr
\theta\buildrel{\tau\to 0}\over\sim&\;\theta_0 + b\tau^\b\,,\cr}\eqno\eq$$
where $\b=1-\a$ (or $\b=1-2\a$ for RN).
There is another possibility, with 
$\a=\b=2$.
Now, from  the energy squared,
$$e^2=\dot r^2 + (r-r_+)(r-r_-)(\dot\theta^2 +
\sin^2\theta)\,,\eqno\eq$$
and the positivity of $(|r_0|-r_+)(|r_0|-r_-)$, we see 
that $\a$ and $\b$ are necessarily greater or equal than
1, thus excluding the generic behaviour $\a+\b=1$ (or $2\a+\b=1$
for RN).
We stress that collapse is indeed possible, as has been
seen in the exact solution \eqsol\ presented above for
$a=M-|r_0|$, but a very fine tuning of the parameters of
the infalling string is required to avoid the repulsion of 
the singularity.

It is important to observe that the most interesting effects observed
in the propagation of classical strings within our approach appear
already for solutions of Einstein-Maxwell's theory, without having to
investigate tree-level string vacua. 
\ack
I.L. Egusquiza is supported by a Basque Government Fellowship.
\refout
\figout
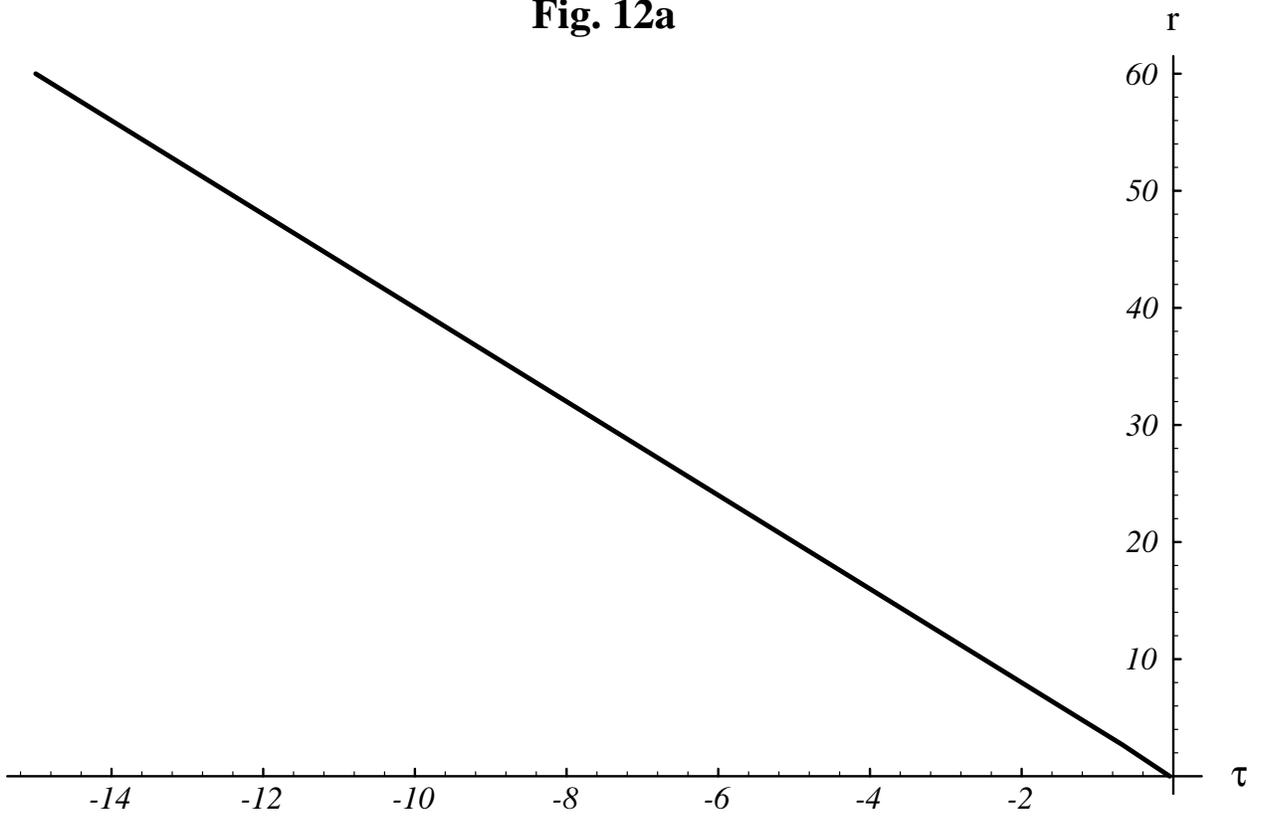
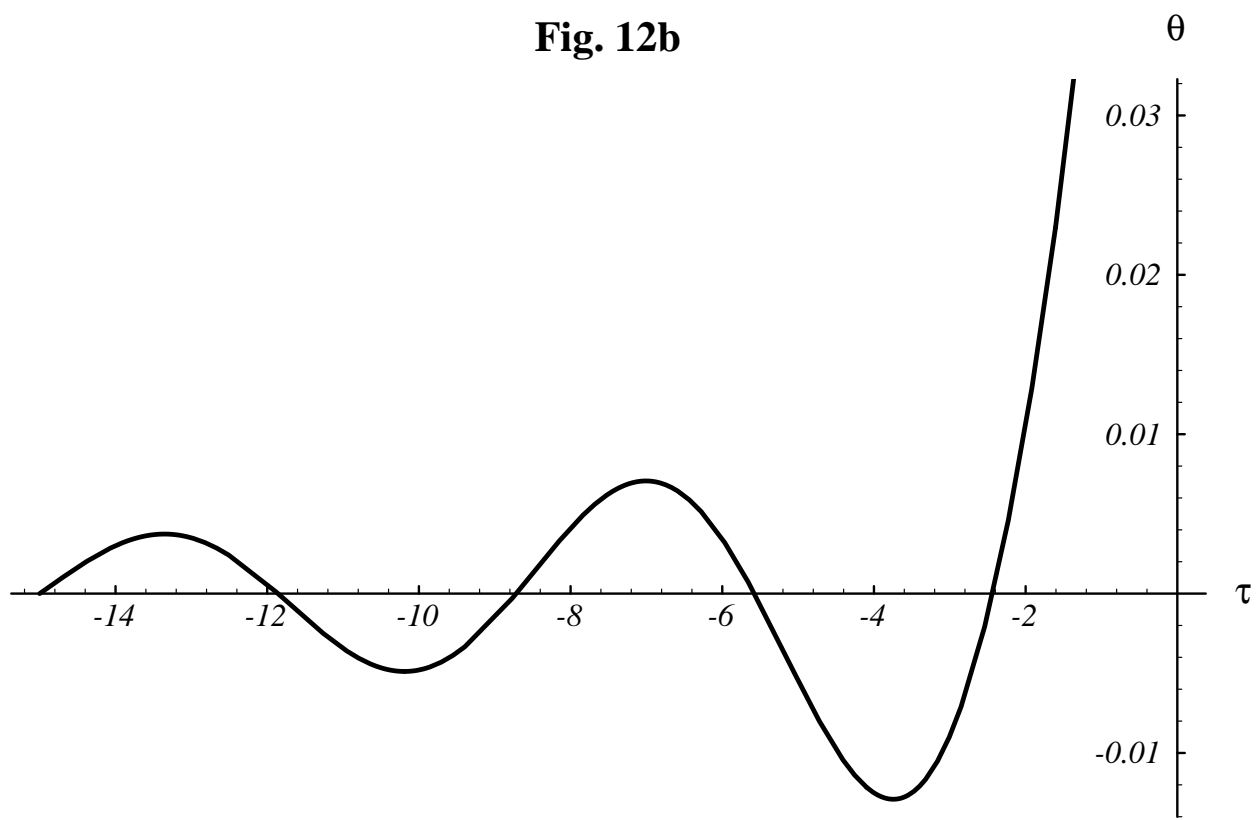
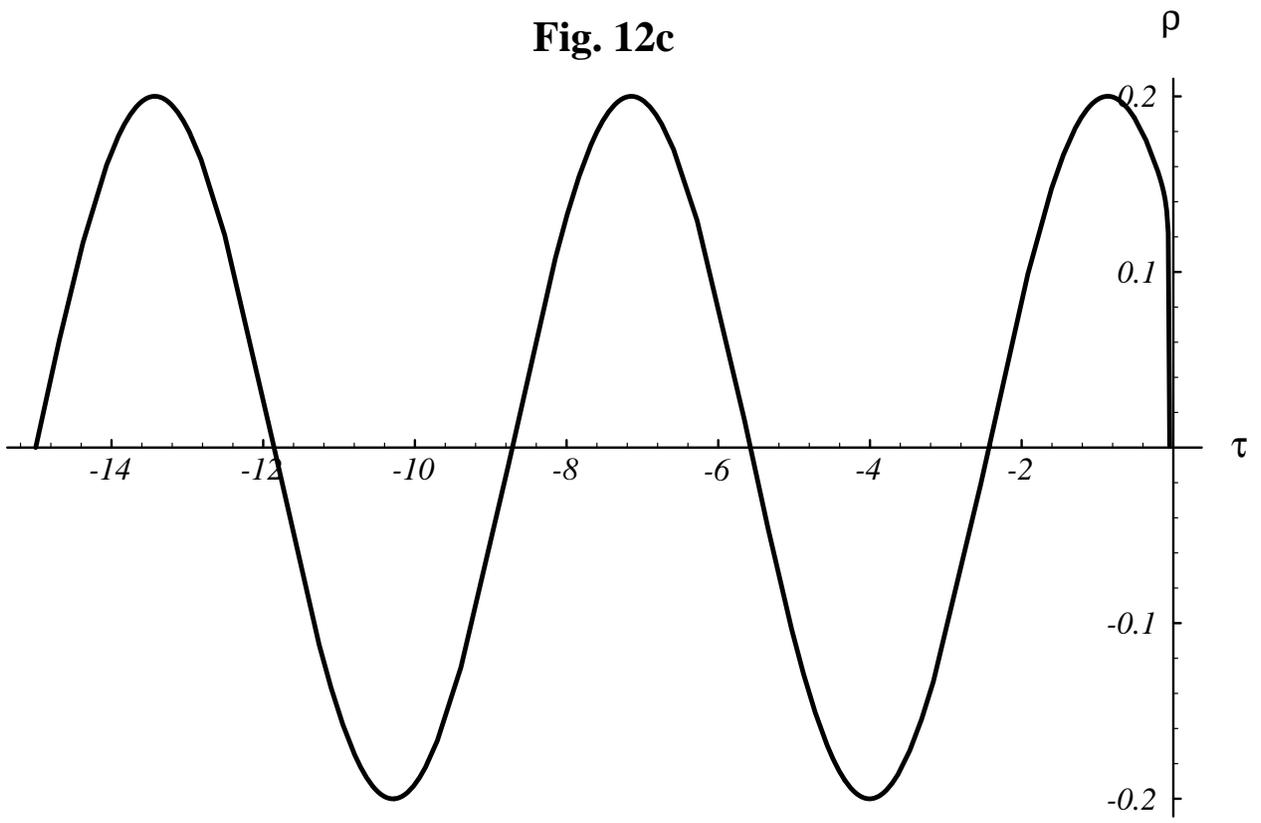
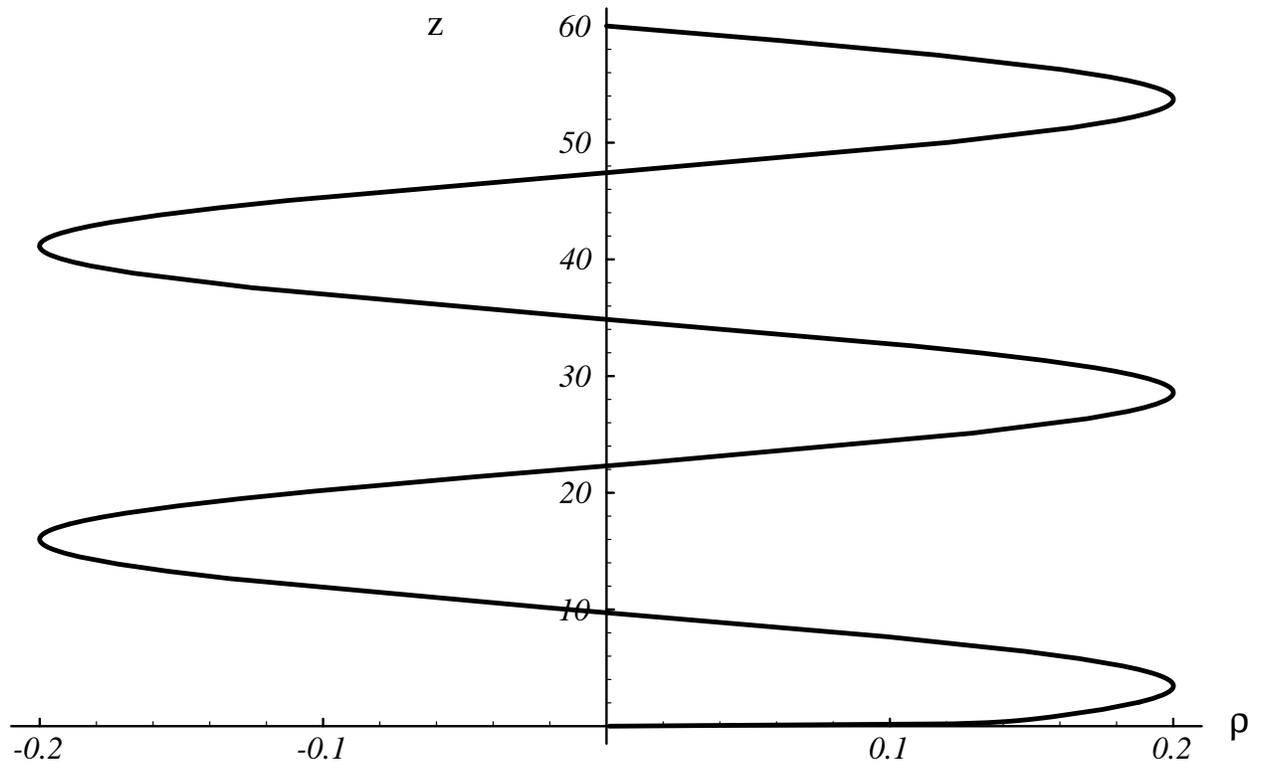
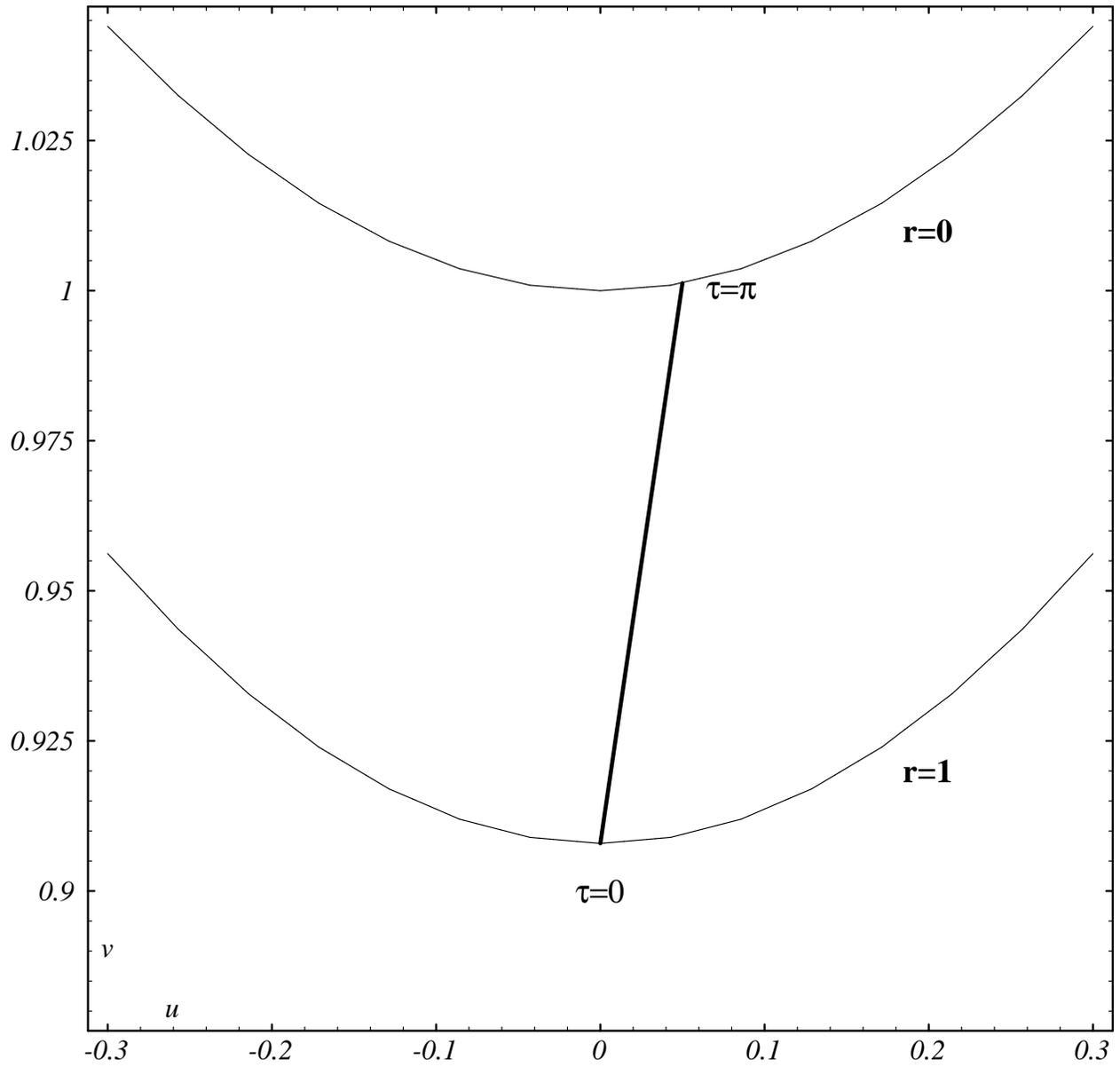
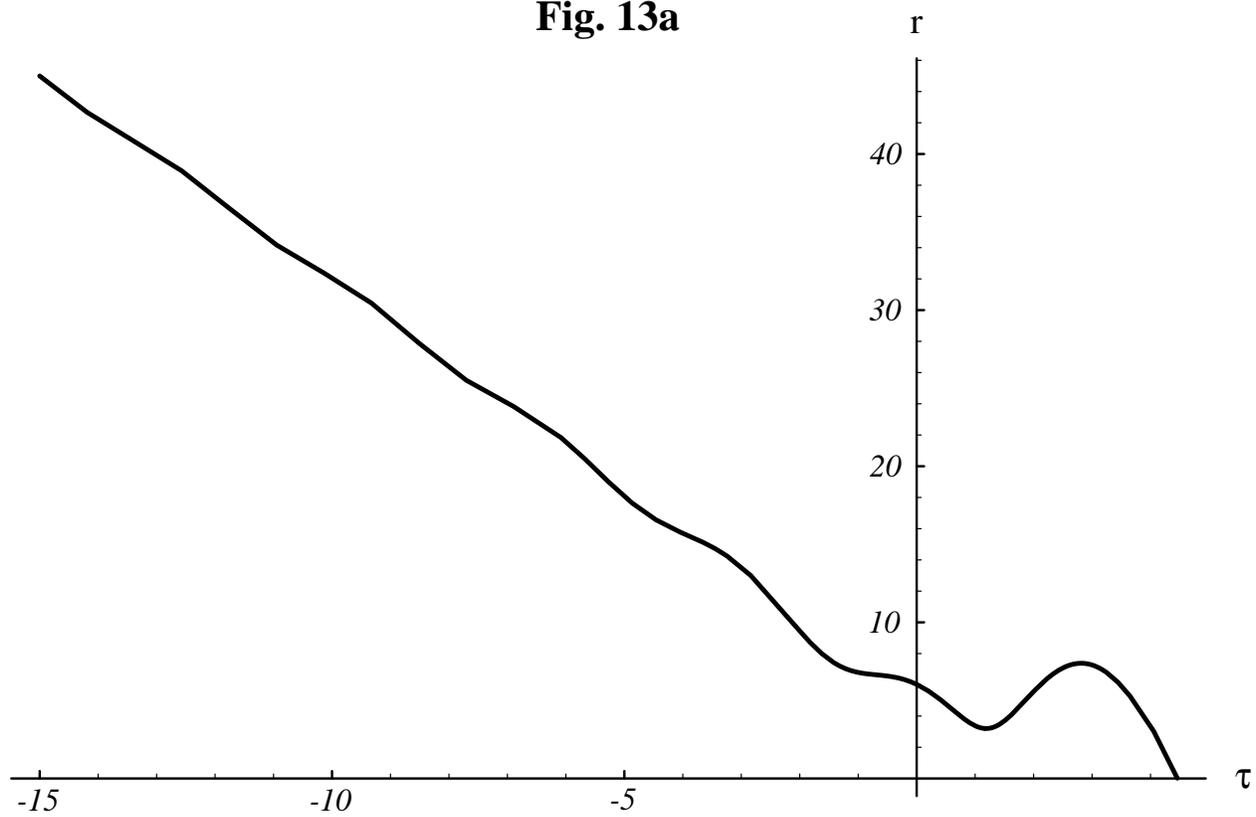
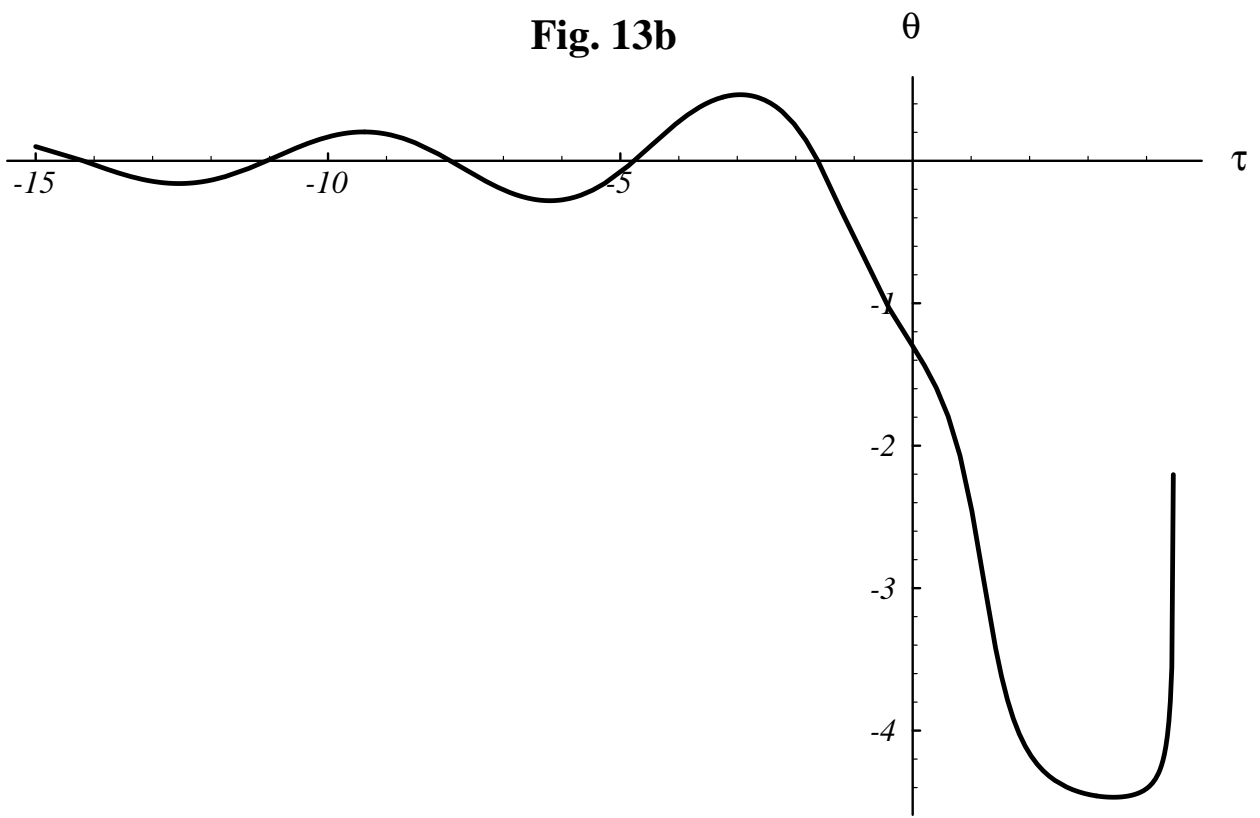
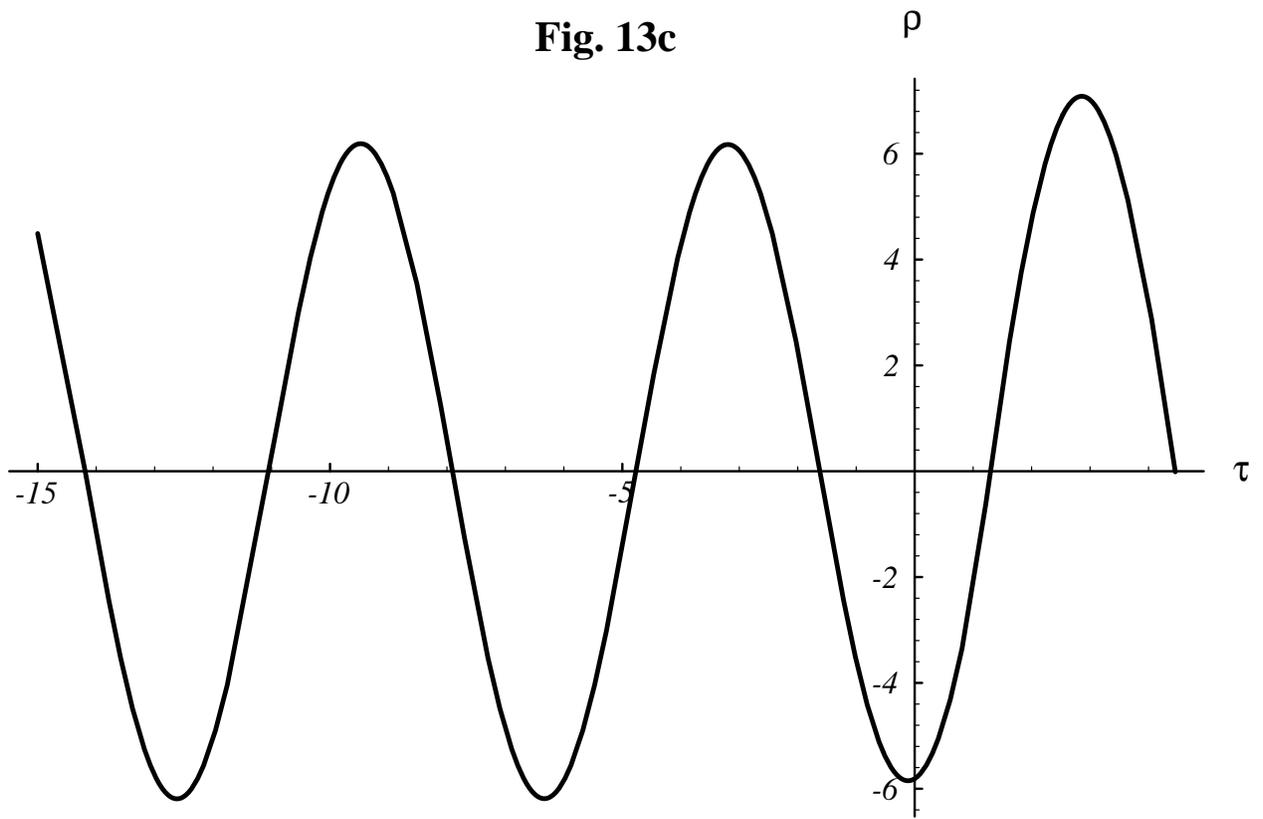
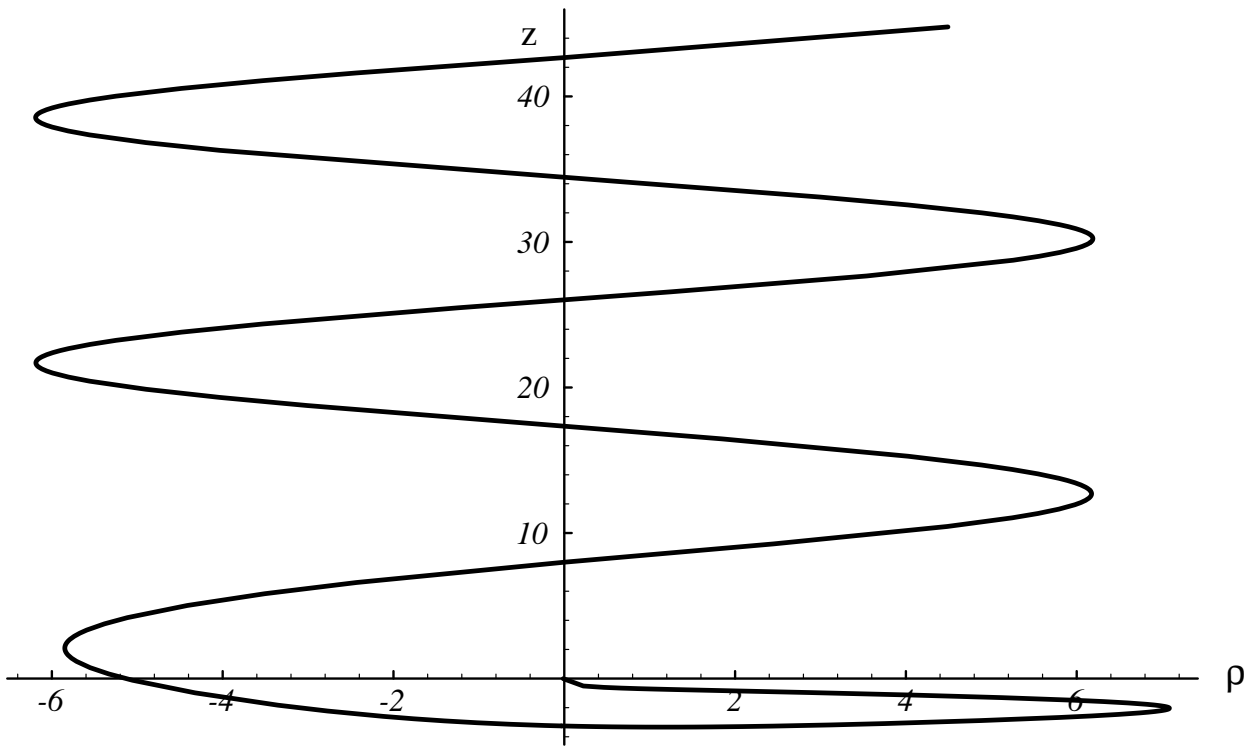
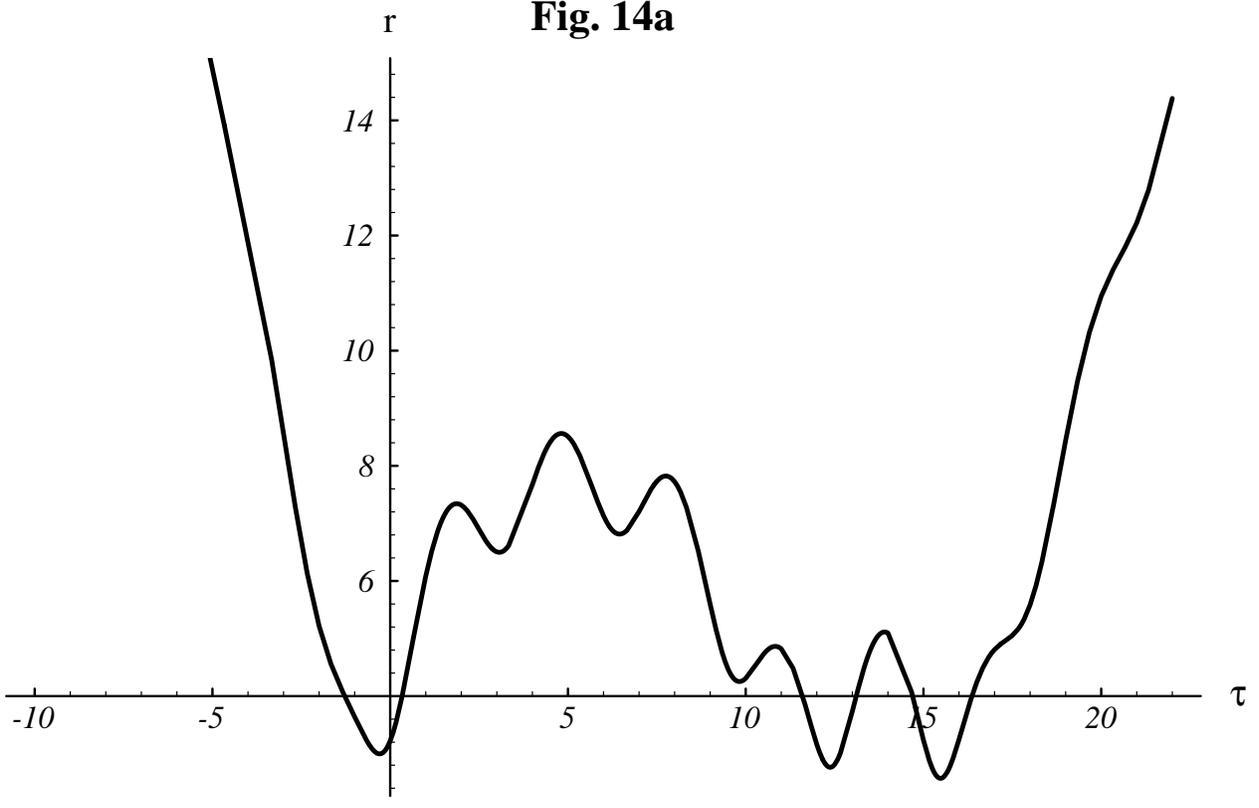
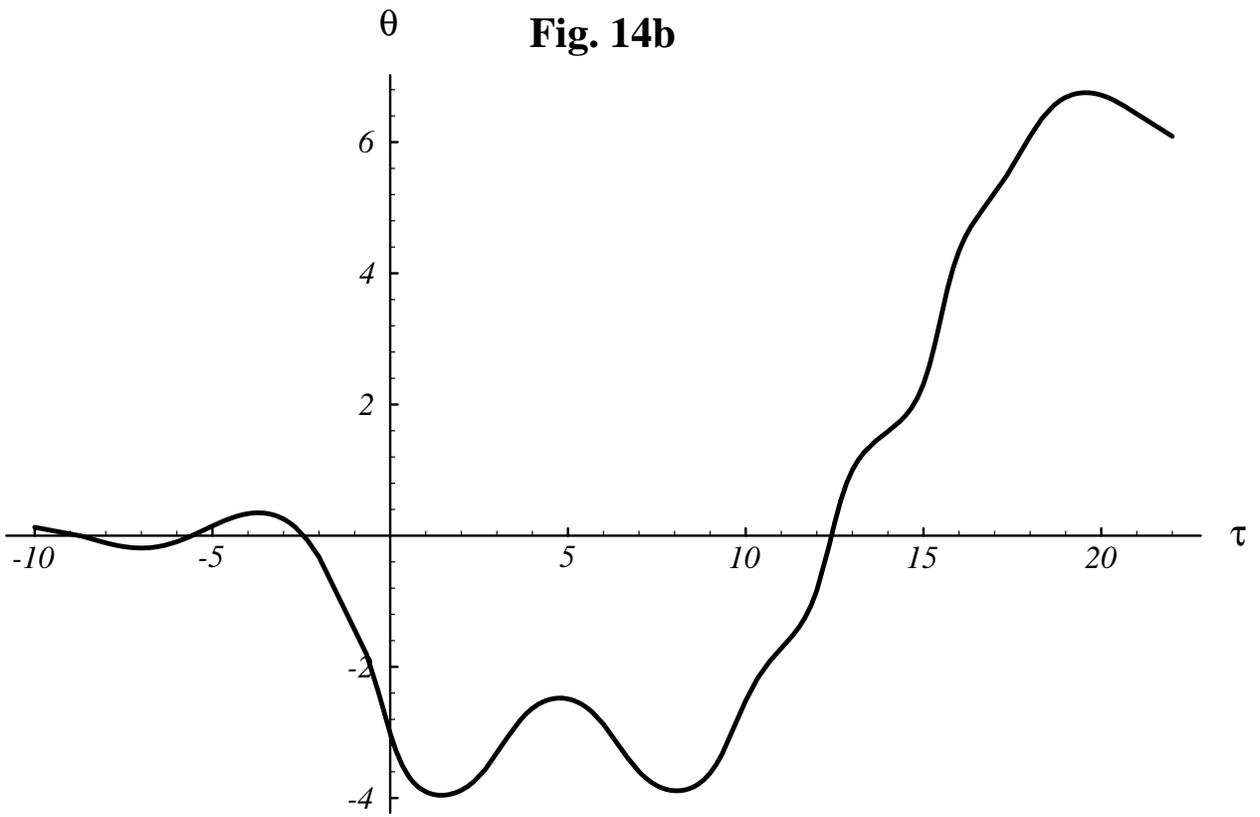
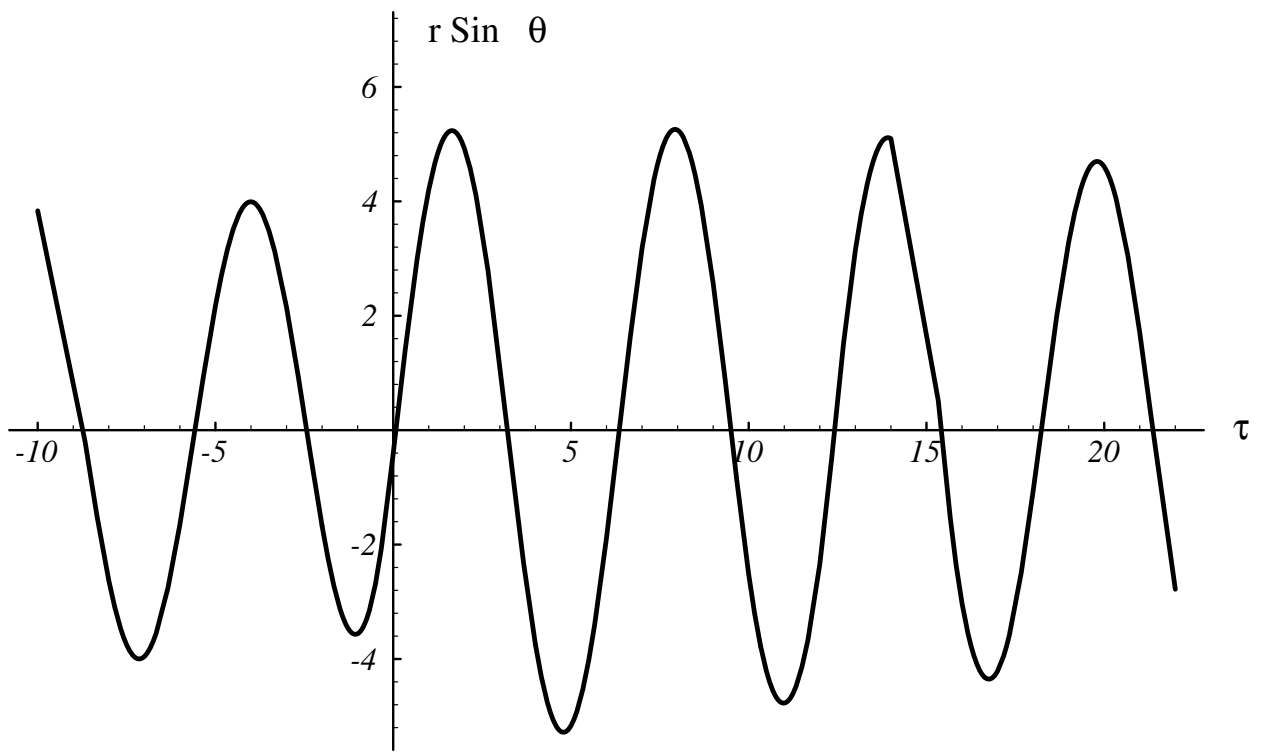
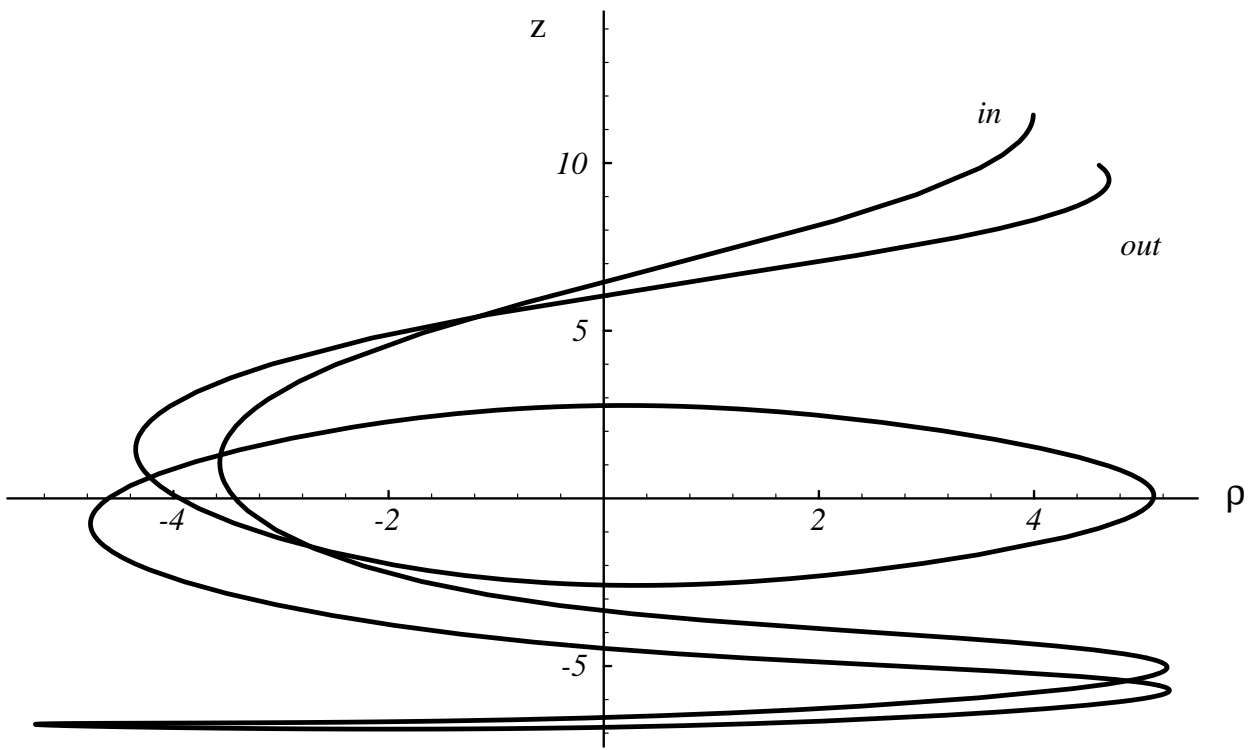
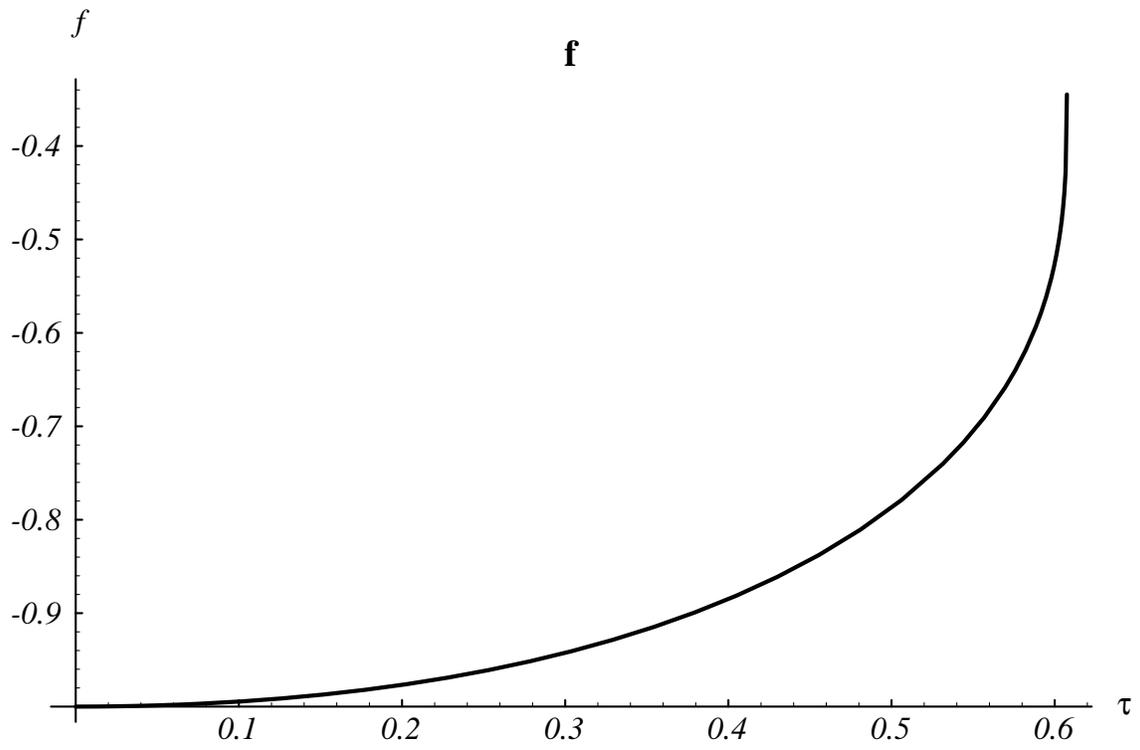
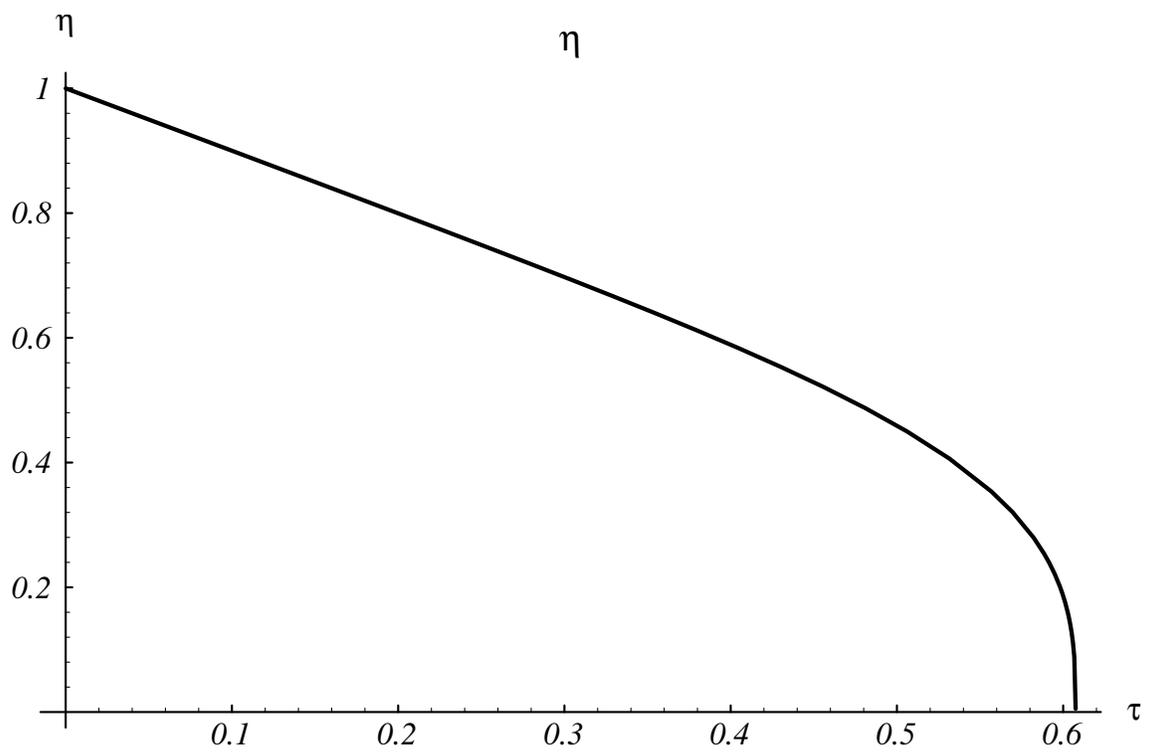
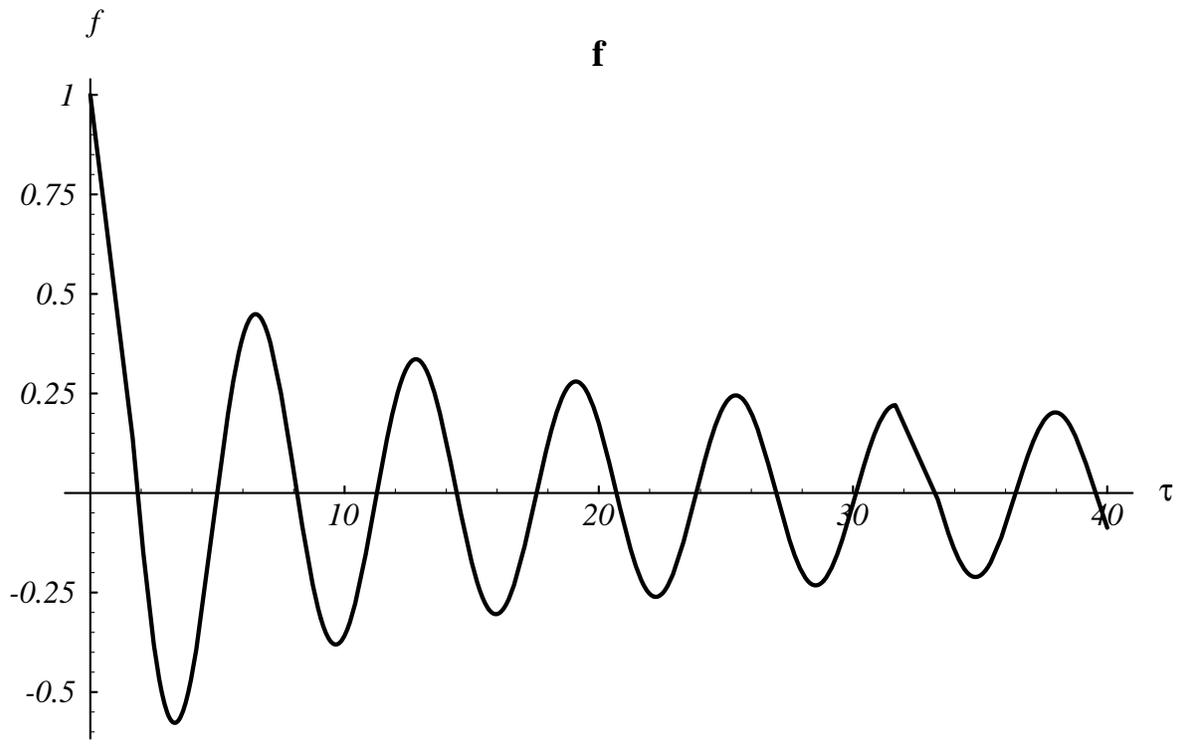
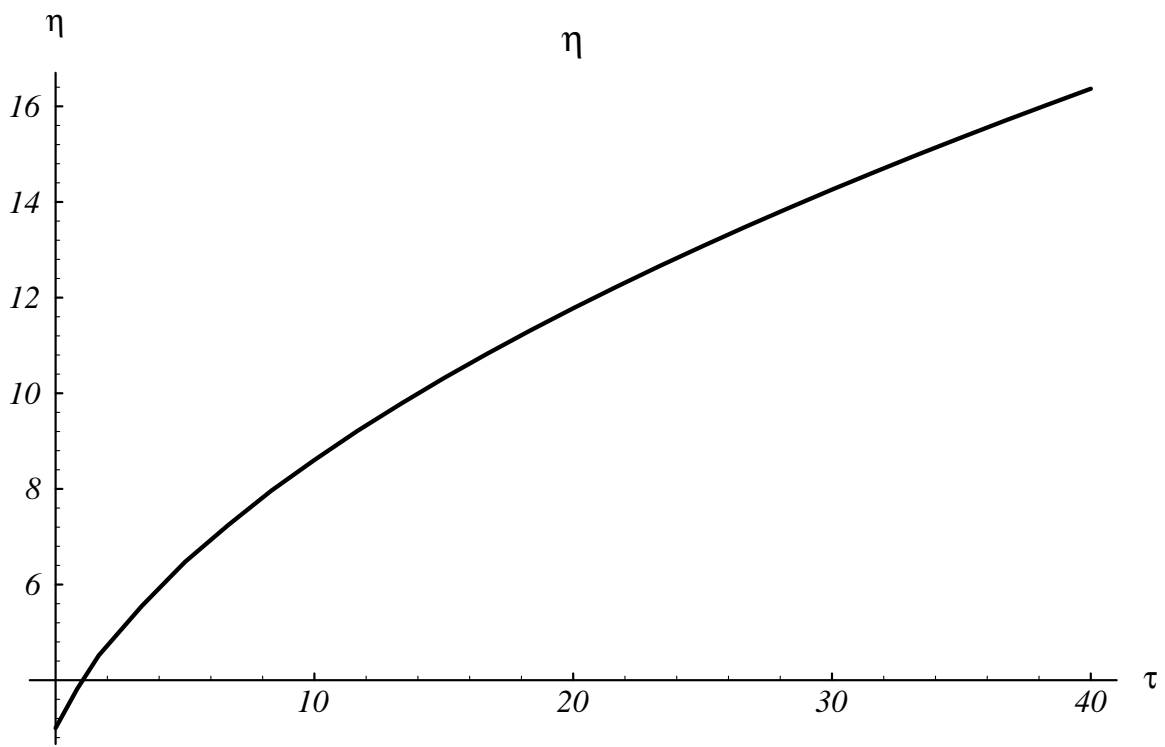
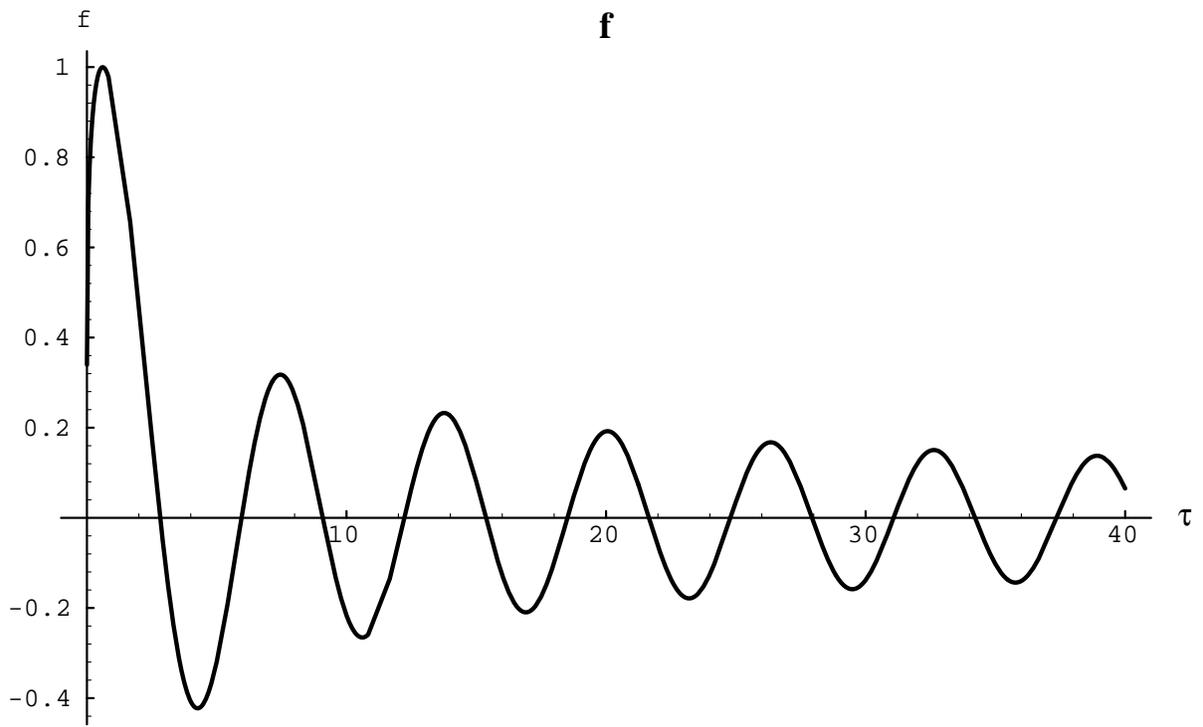
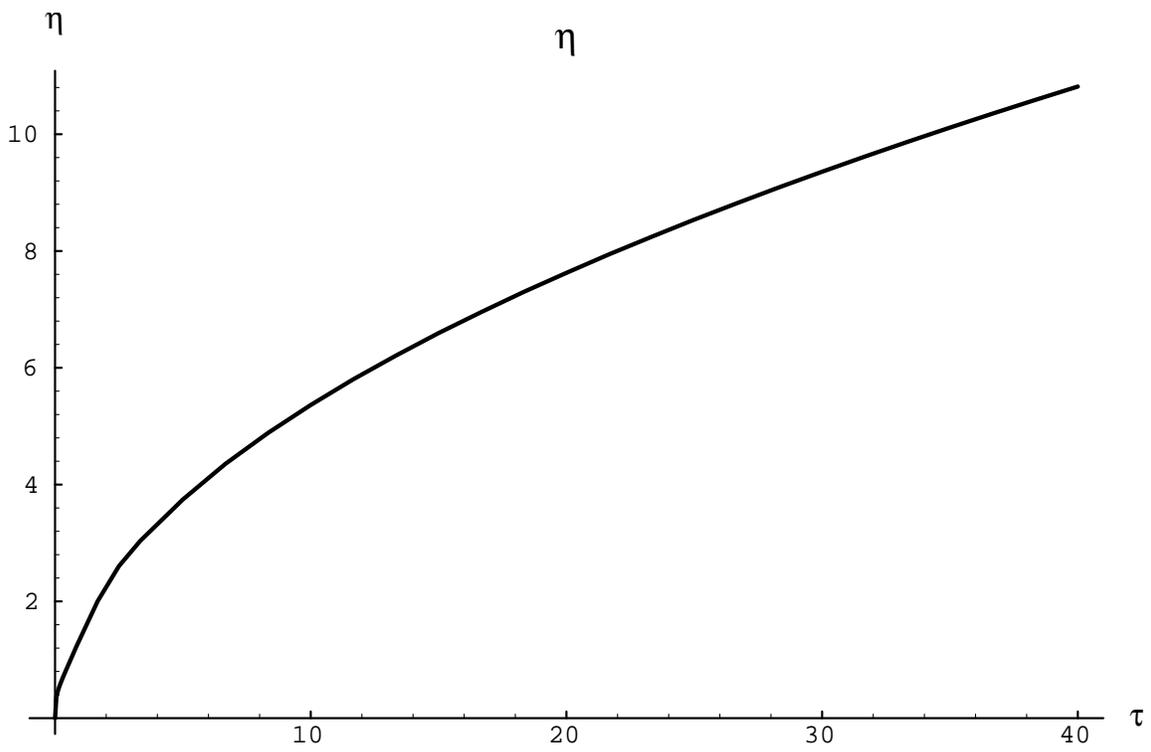
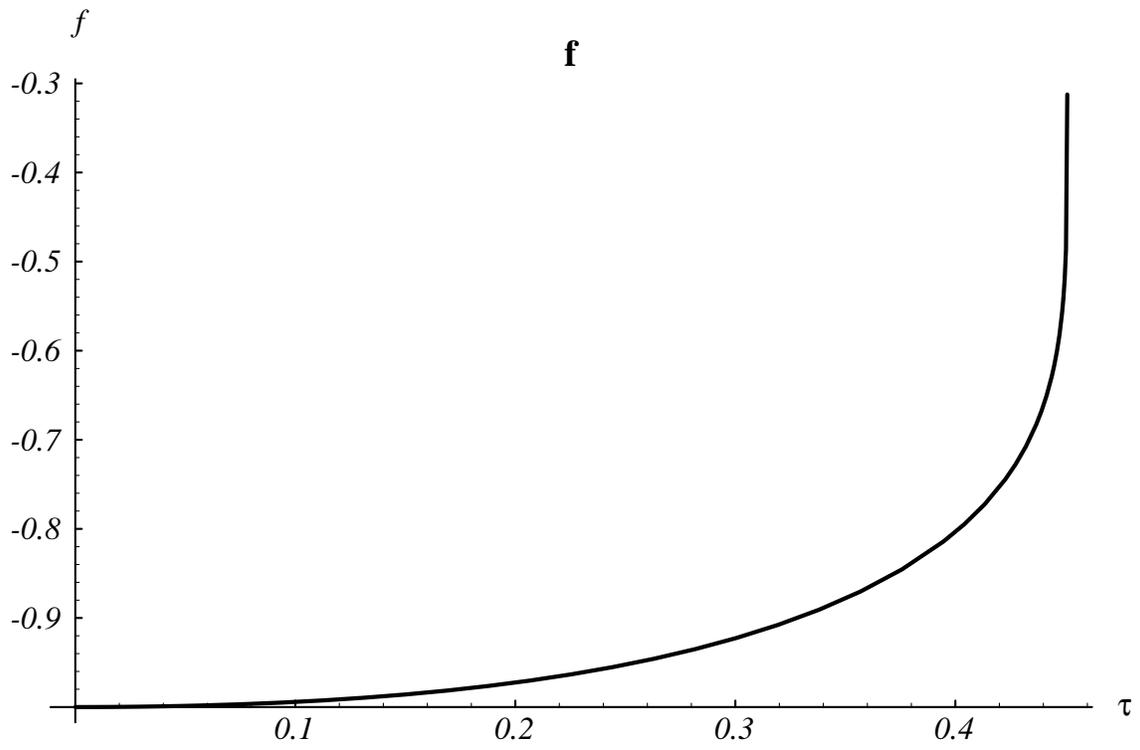
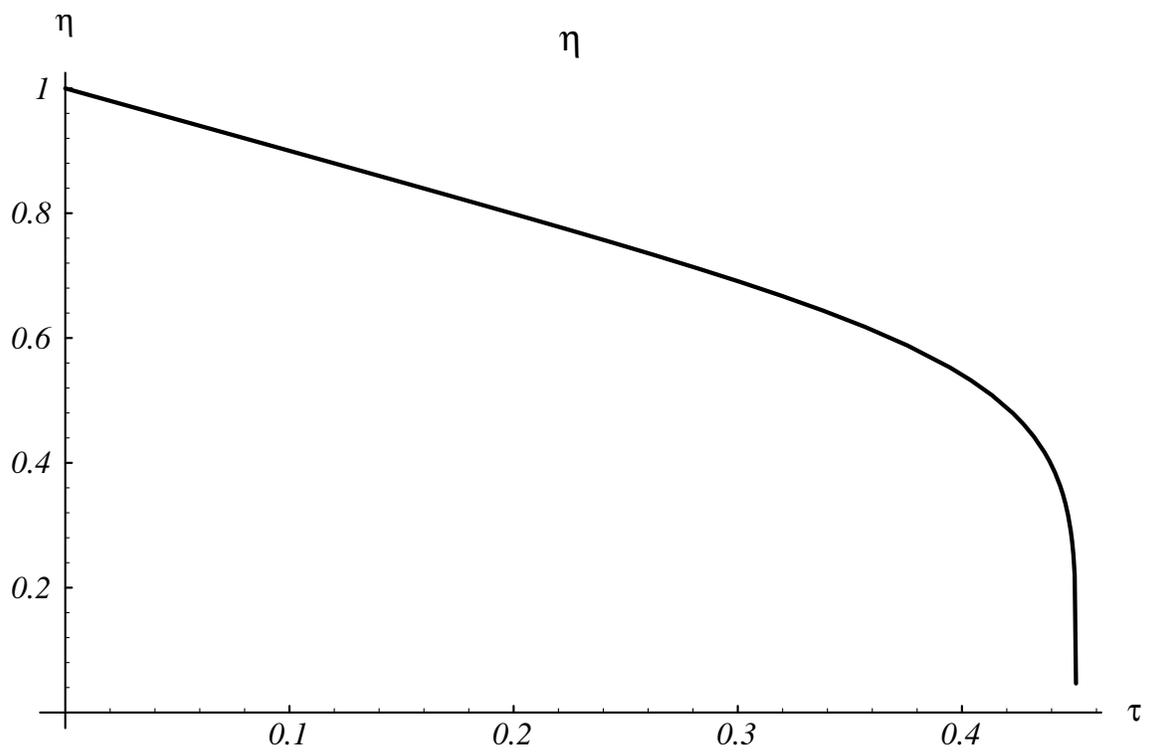
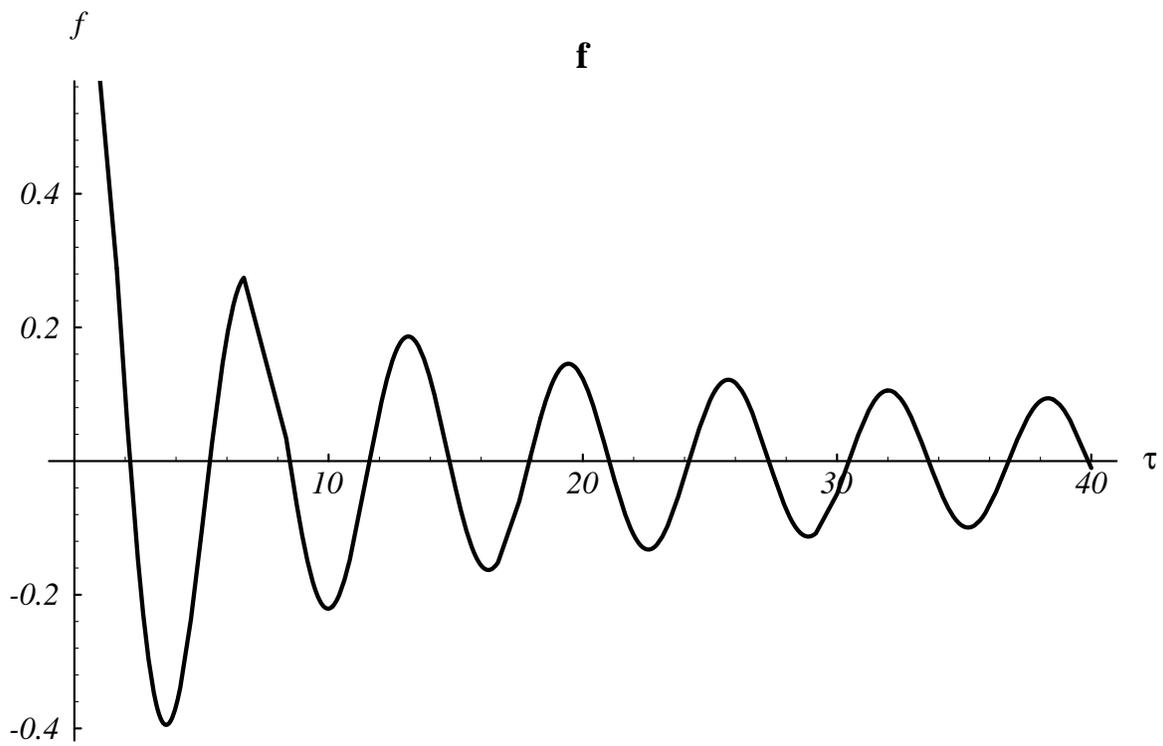
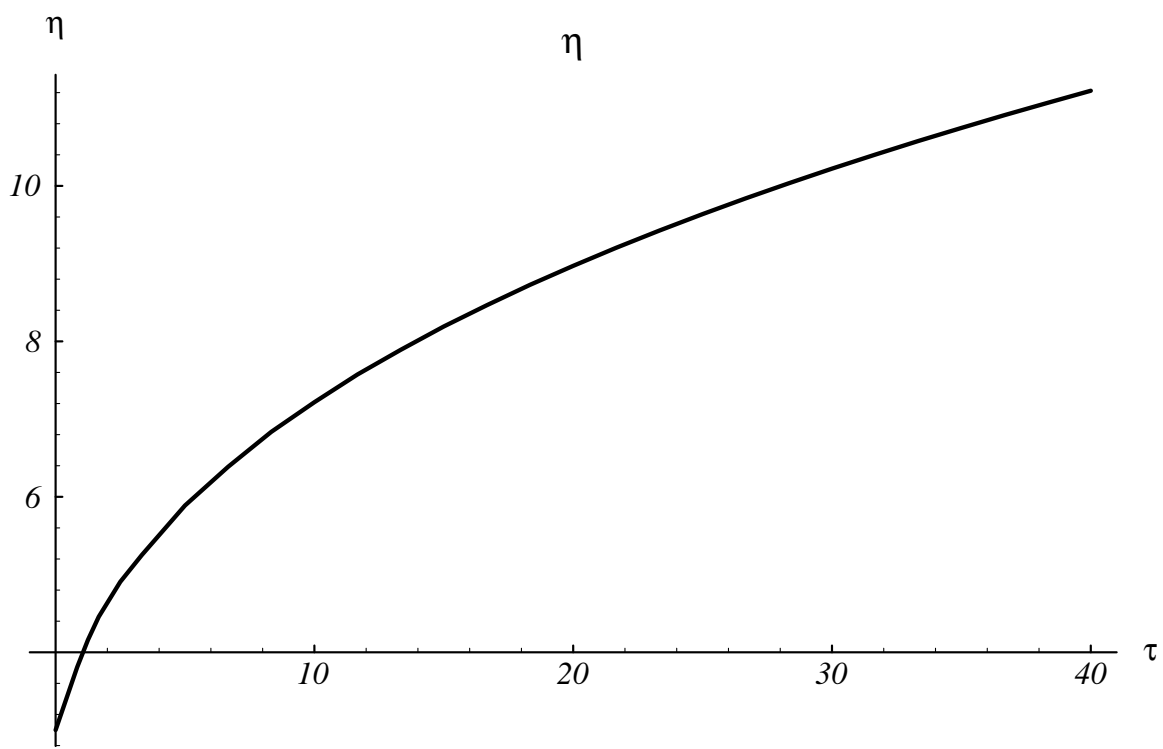
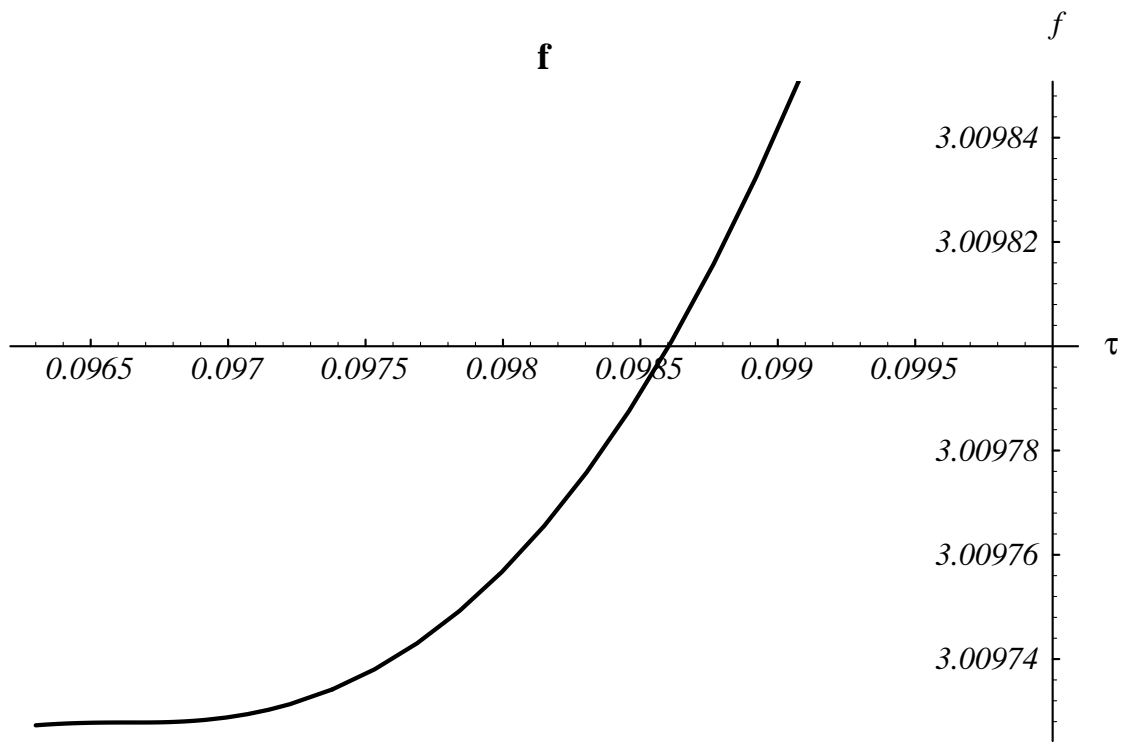
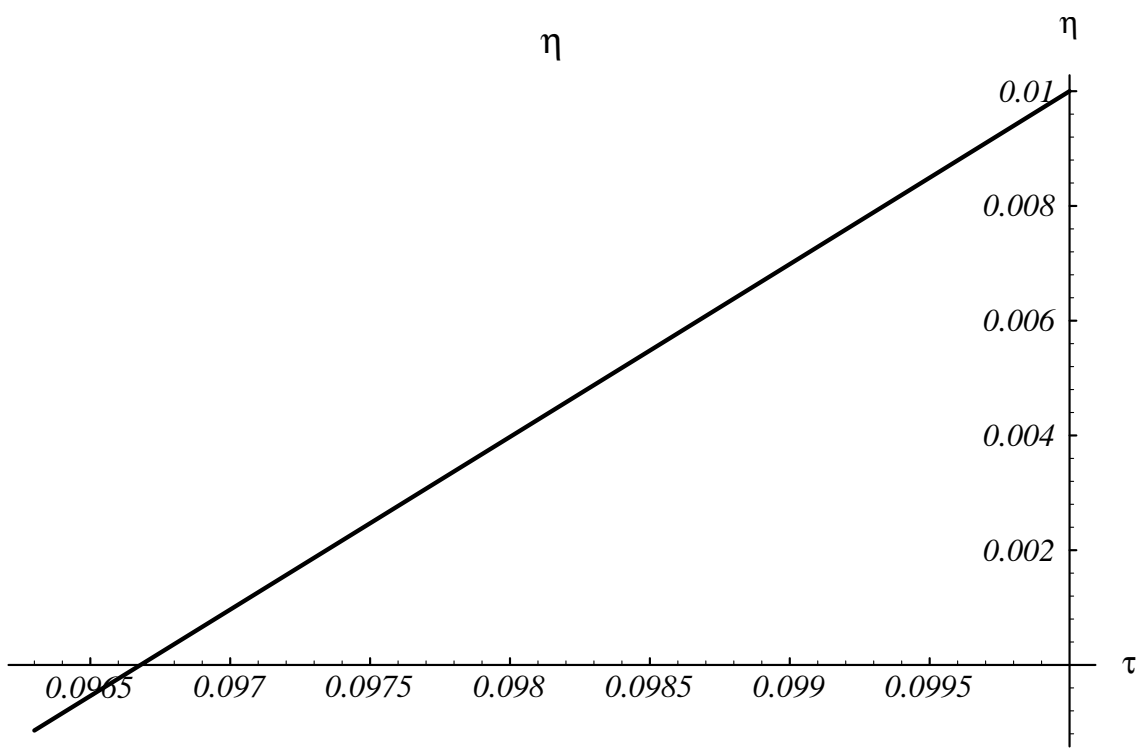
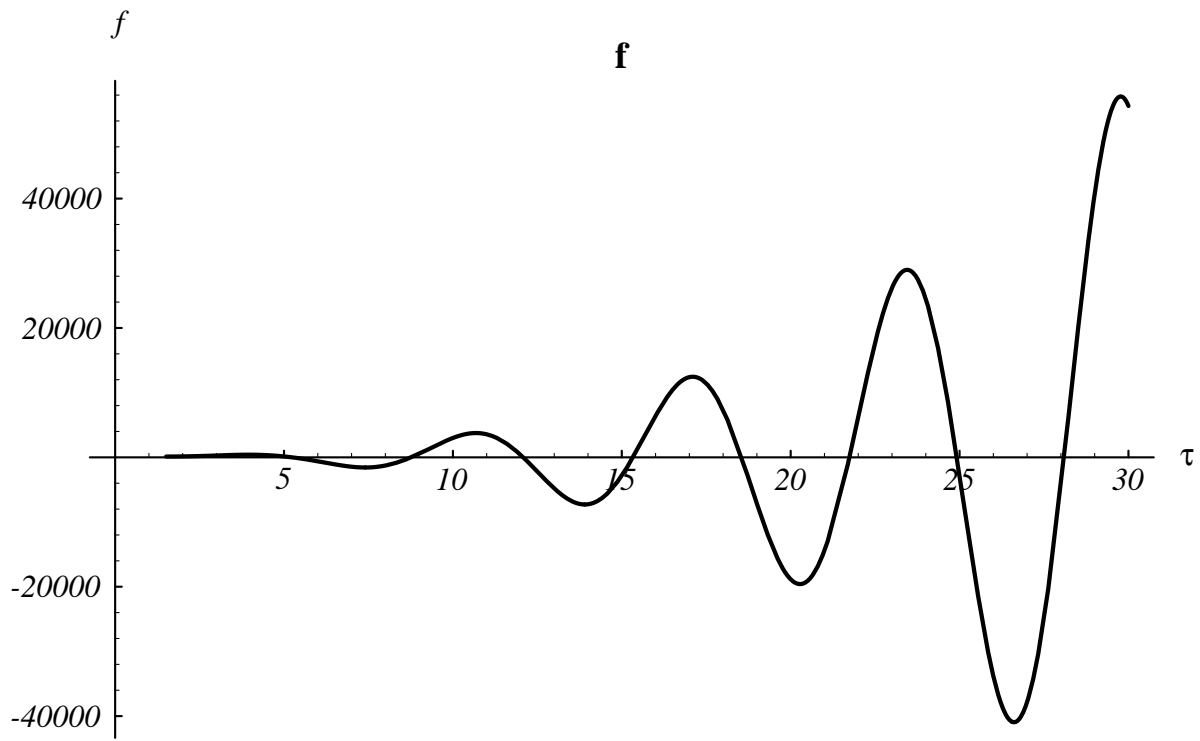
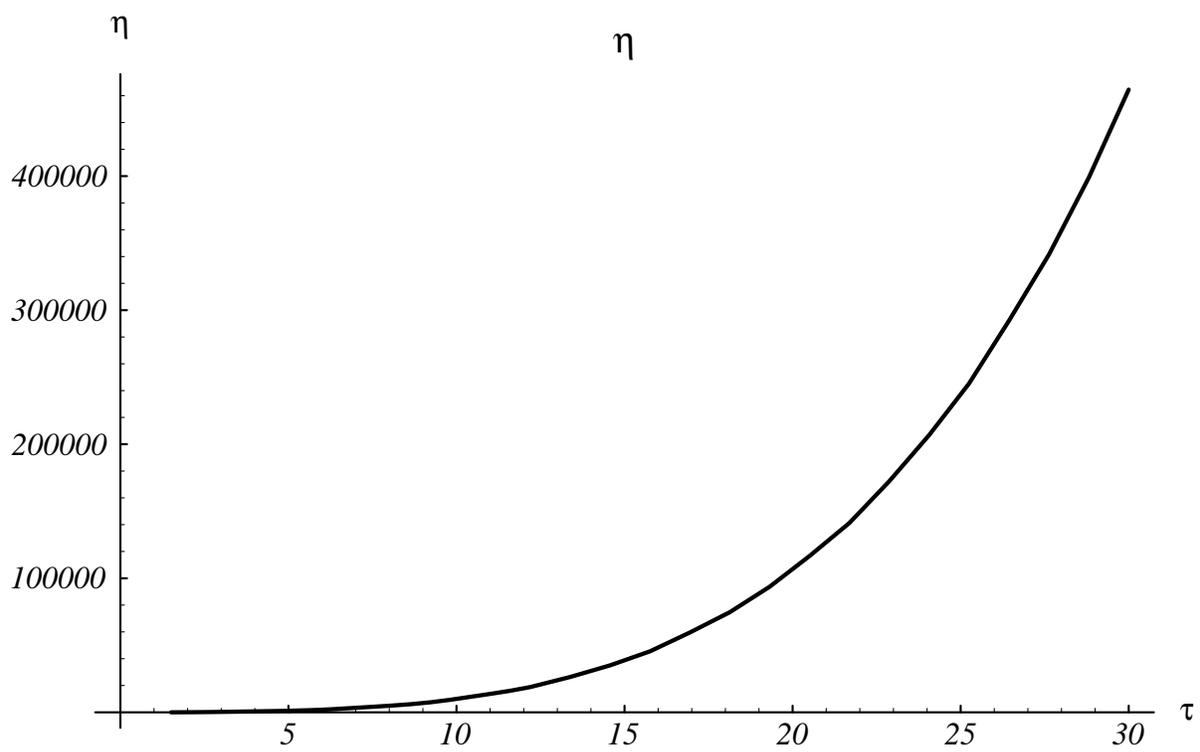
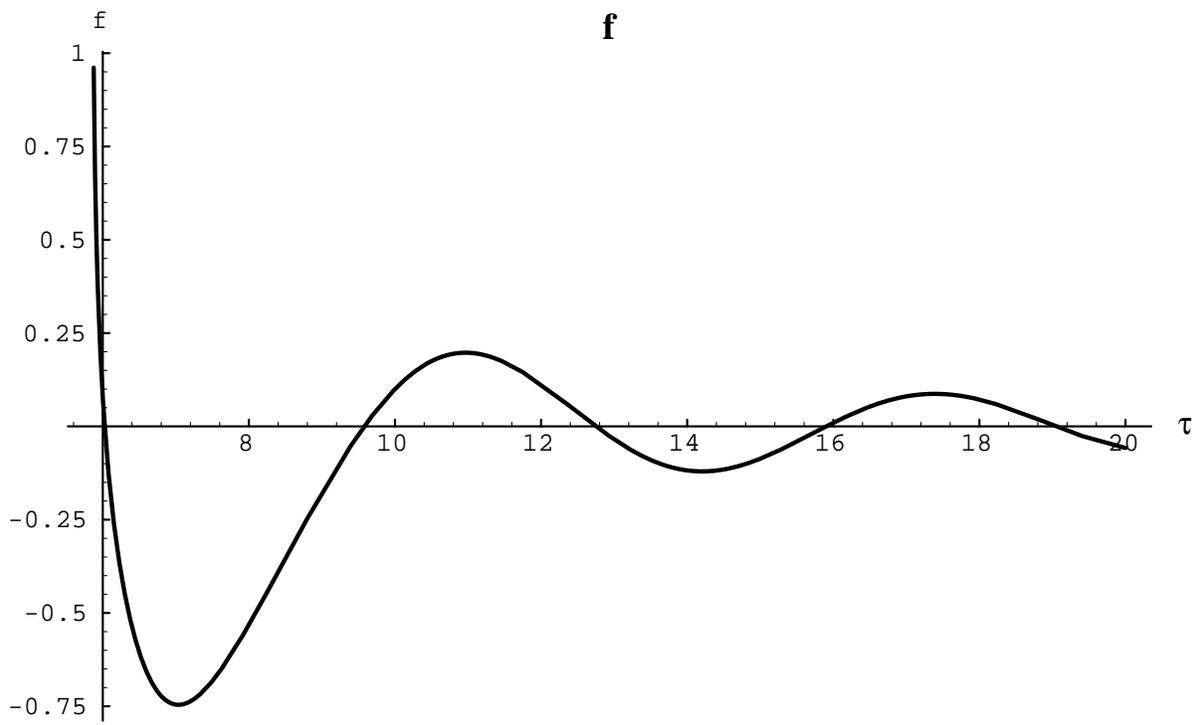
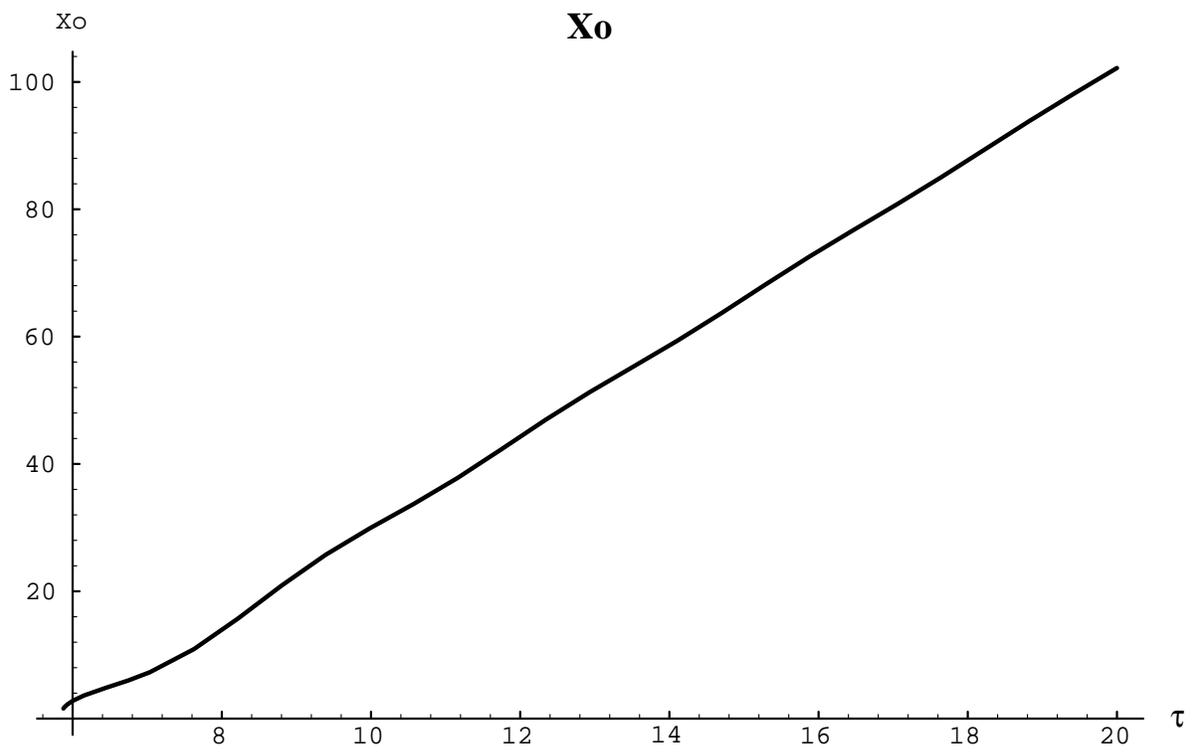
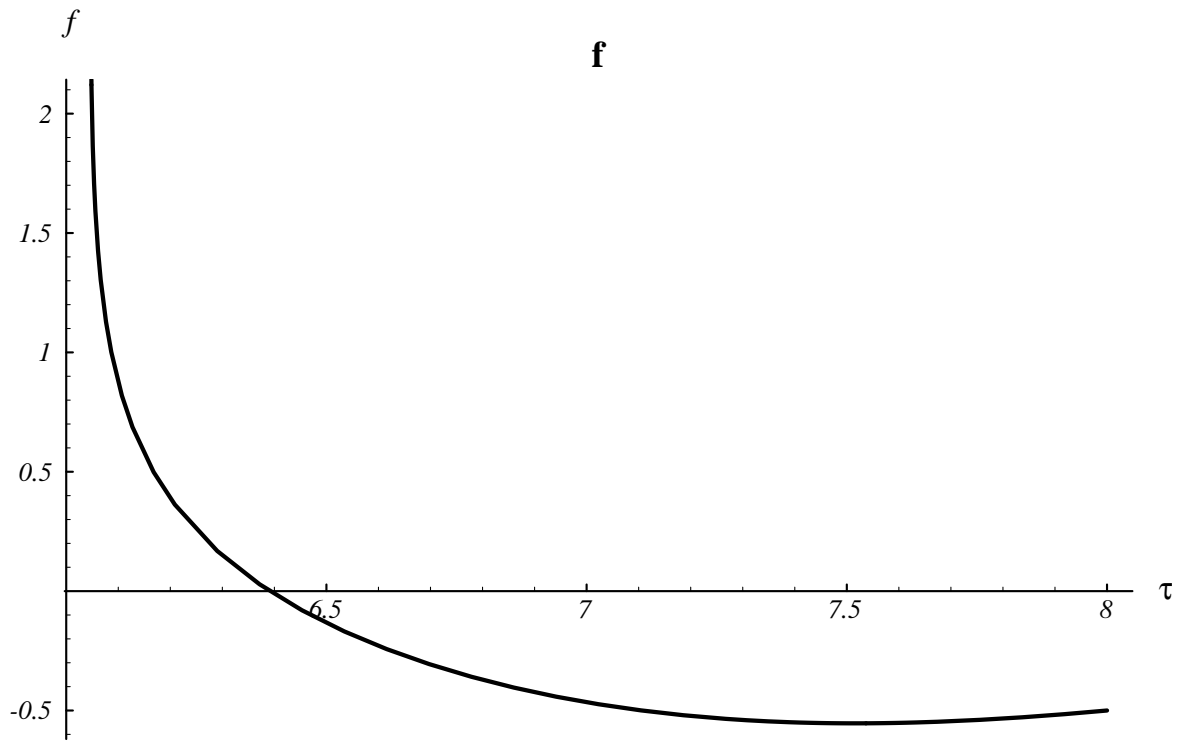
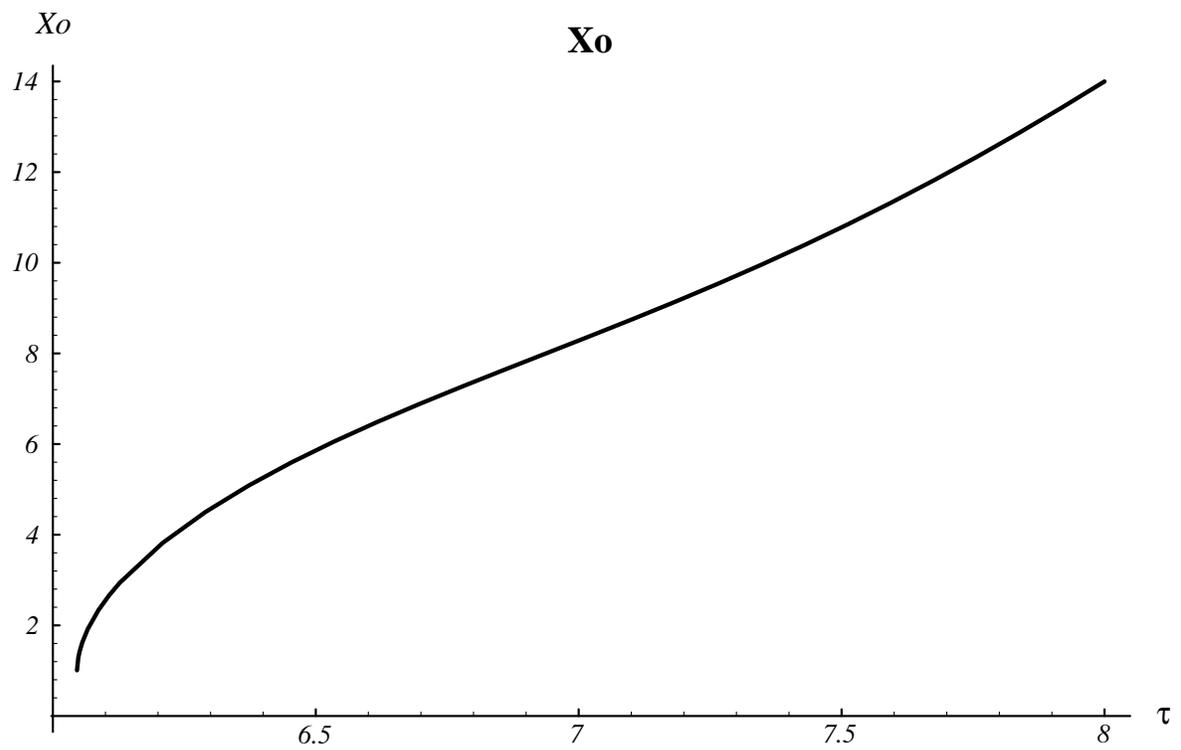
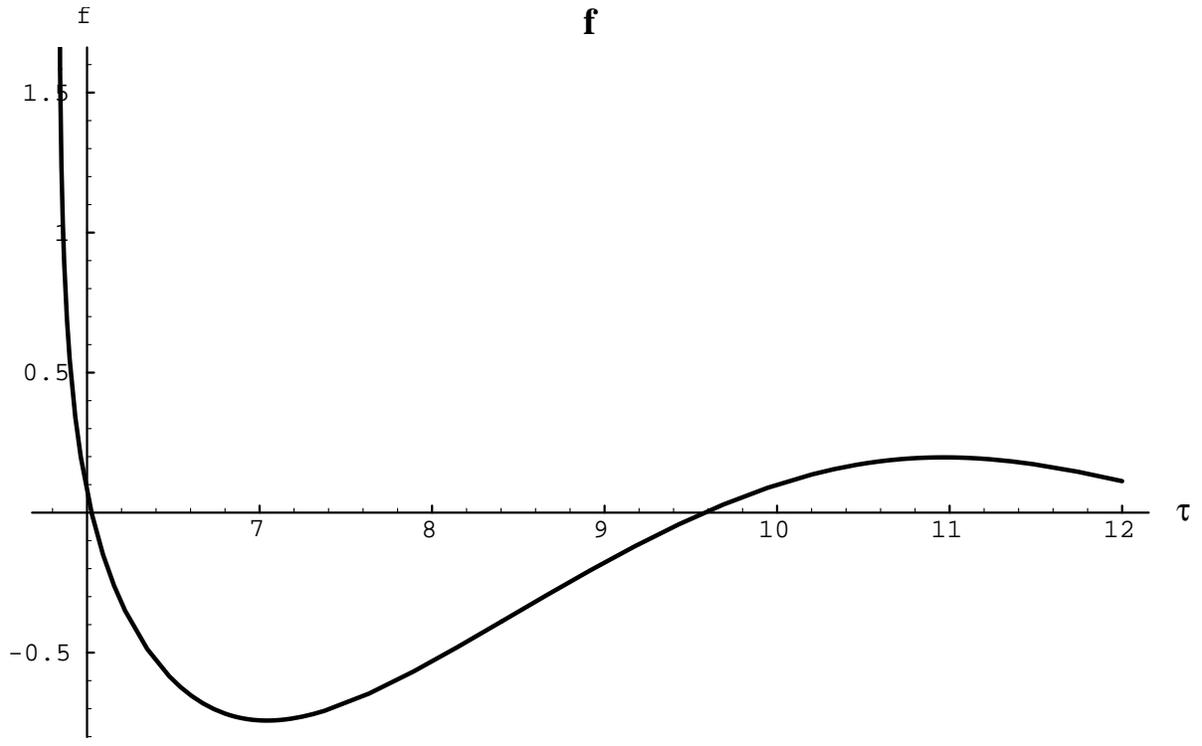
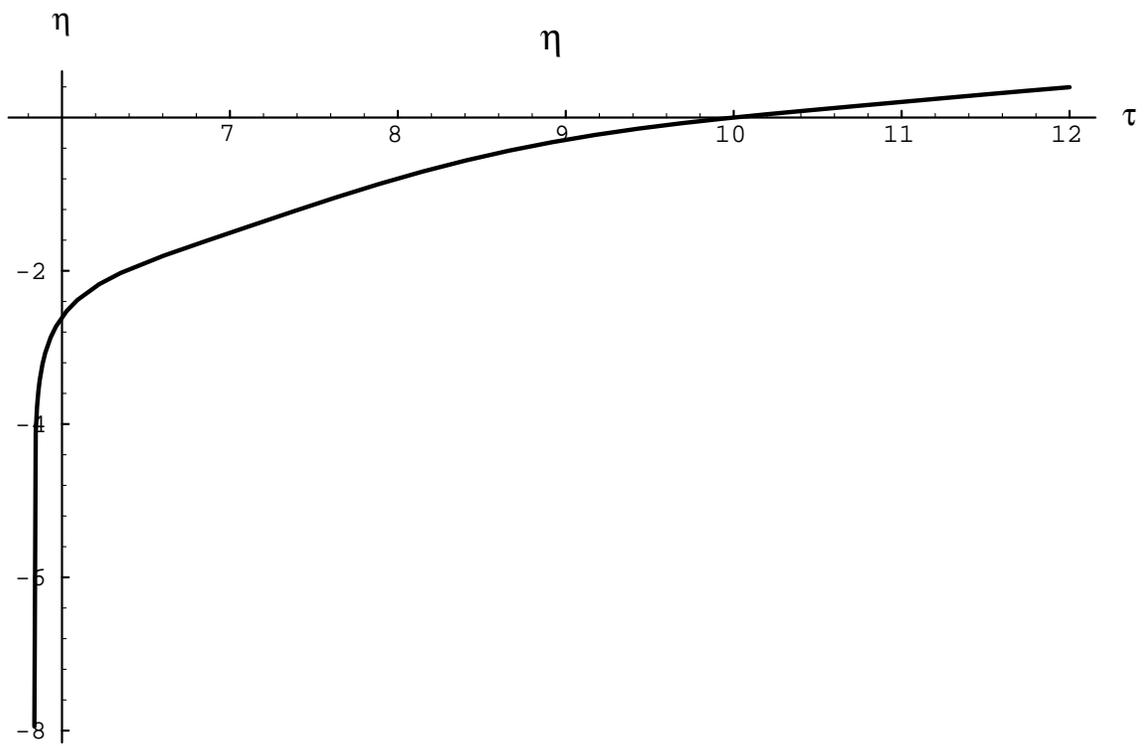
\bye